\documentclass[iop]{emulateapj}
\usepackage{graphicx}
\usepackage{amsmath}
\usepackage{amssymb}
\usepackage{here}
\usepackage{multirow}
\usepackage{rotating}
\usepackage{booktabs}
\usepackage{times}

\newcommand{\zmag}{\ensuremath{z^\prime}}
\newcommand{\Msun}{\ensuremath{M_{\odot}}}

\newcommand{\ar}{\arcsec}
\newcommand{\wz}{\ensuremath{z_{850}}}

\newcommand{\mg}{\mu_\gamma}
\newcommand{\s}{\sigma}
\newcommand{\sg}{\sigma_\gamma}

\slugcomment{Accepted for Publication in ApJ}

\shorttitle{Intrinsic Shape of sBzK in GOODS-N}
\shortauthors{Yuma et al.}

\begin{document}

\title{Intrinsic Shape of Star-Forming BzK Galaxies at $z\sim2$ in GOODS-N}
\author{Suraphong Yuma        \altaffilmark{1}, 
        Kouji Ohta        \altaffilmark{1},
        Kiyoto Yabe        \altaffilmark{1},	
        Masaru Kajisawa 	\altaffilmark{2},
        Takashi Ichikawa 	\altaffilmark{3}
        }

\email{yuma@kusastro.kyoto-u.ac.jp}

\begin{abstract}
We study structure of star-forming galaxies at $z\sim2$ in GOODS-N field 
selected %based on the $B-z$ and $z-K$ color diagram 
as sBzK galaxies down to $K_{\rm AB} <24.0$ mag. 
Among 1029 sBzK galaxies, 551 galaxies 
(54\%) show a single component in ACS/F850LP image 
obtained with the Hubble Space Telescope; 
the rest show multiple components. % suggesting 
%that they are interacting or merging systems. 
We fit the single-component sBzK galaxies 
with the single S\'ersic profile using the ACS/F850LP image 
and find that a majority of them (64\%) show S\'ersic index of 
$n=0.5-2.5$, indicating that they have a disk-like structure. 
The resulting effective radii typically range from 1.0 to 3.0 kpc in 
the rest-frame UV wavelength. 
After correcting the effective radii to those in the rest-frame optical wavelength, 
we find that the single-component sBzK galaxies locate in the region where 
the local and $z\sim1$ disk galaxies distribute in the stellar mass-size 
diagram, suggesting comparable surface 
stellar mass density between the sBzK and $z\sim0-1$ disk galaxies. 
All these properties suggest that the single-component sBzK galaxies 
are progenitors of the present-day disk galaxies. 
However, by studying their intrinsic shape through comparison between 
the observed distribution of apparent axial ratios and the distribution for 
triaxial models with axes ($A>B>C$), we find that the mean $B/A$ 
ratio is $0.61^{+0.05}_{-0.08}$ and disk thickness $C/A$ is $0.28^{+0.03}_{-0.04}$. 
This indicates that the single-component sBzK galaxies at $z\sim2$ have a bar-like 
or oval shape rather than a round disk shape. 
The shape seems to resemble to a bar/oval structure that 
form through bar instability; if it is the case, the intrinsic shape 
may give us a clue to understand dynamical 
evolution of baryonic matter in a dark matter halo. 

\end{abstract}

\keywords{galaxies: evolution --- galaxies: formation --- galaxies: high-redshift --- galaxies: structure
        }

\altaffiltext{1}{Department of Astronomy, Kyoto University, Sakyo-ku, Kyoto 606-8502, Japan}
\altaffiltext{2}{Research Center for Space and Cosmic Evolution, Ehime University, Bunkyo-cho, Matsuyama, 790-8577, Japan}
\altaffiltext{3}{Astronomical Institute, Tohoku University, Aoba, Sendai 980-8578, Japan}

%----------------------Main text--------------------------
\section{Introduction}\label{sec:intro}

Formation of disk galaxies is one of the important problems 
in astronomy. 
Many theoretical studies proposed scenarios of the disk formation 
through analytical approaches \citep[e.g.,][]{white78, fall80} or simulations 
\citep[e.g.,][]{navarro91, navarro94, navarro97, barnes02, springel05, 
robertson06, sales10, dutton11}. 
However, formation epoch of disks and high-redshift counterparts of the
present-day disk galaxies are still not clear. 
Observational studies have been revealing that disk galaxies 
are already in place at $z\sim 1$. 
Although a mild luminosity evolution ($\sim 1$ mag) can be
seen up to $z \sim1$ \citep[e.g.,][]{brinch98,sca}, 
stellar mass-size relation does not change significantly \citep{barden05}.
Stellar-mass function of disk galaxies also does not evolve so 
much up to $z\sim1$ in the massive part 
\citep[e.g.,][]{bundy05, pannella06}.
Furthermore, disk-size (scale length for the exponential law)  function  does not 
show significant evolution for the  larger disks \citep{lilly98, sargent07}. 
These suggest that the formation of disk galaxies is beyond $z\sim1$. 
Meanwhile,  no clear counterparts of disk galaxies are known at $z>3$. 
Some of Lyman break galaxies (LBGs) at $z\sim3$ show the 
surface brightness distribution with S\'ersic index of $n\sim1$ 
\citep[e.g.,][]{steidel96}. 
\cite{akiyama08} found that most of their LBG sample at $z\sim3$ 
show  S\'ersic profiles with $ 1 < n < 2$ in the rest-frame optical 
wavelength by AO-assisted $K-$band imaging observations. 
They showed, however, that the stellar-mass density is too high to evolve into
the present-day disk galaxies, suggesting that the LBGs are progenitors
of elliptical galaxies.
Furthermore, clustering amplitude of LBGs  is very large indicating that 
they reside in massive dark halos and their destination would be 
giant elliptical galaxies in rich clusters 
\citep[e.g.,][]{ouchi01,gm01}.
Thus LBGs (at least bright/massive LBGs) are presumably not a
direct progenitor of present-day disk galaxies.
In other words, no obvious disk galaxy existed at $ z > 3$. 
In fact, \cite{kajisawa01} studied a nearly complete sample of galaxies 
to $z=2$ and found that the Hubble sequence was established 
during $z\sim 1-2$ epoch, though the sample size is not large. 

Therefore, the epoch of $z\sim2$ is important for the understanding
of disk formation. 
The redshift range is, however, so-called redshift desert, 
and isolating galaxies at the epoch is not easy. 
\cite{daddi04} proposed a two-color selection method to isolate 
galaxies at $1.4\lesssim z\lesssim 2.5$ by 
using $B-\zmag$ and $\zmag-K$ colors. 
Galaxies selected by this method are called BzK galaxies.  
Star-forming BzK (sBzK) and passive BzK (pBzK) are defined 
according to the location of a galaxy in the $B-\zmag$ and 
$\zmag-K$ diagram \citep{daddi04}. 
Properties of BzK galaxies have been studies 
\citep[e.g.,][]{daddi04, daddi05, kong06, hayashi07, 
hartley08, yoshikawa10}. 
Among them, \cite{hayashi07} studied the clustering properties of 
faint sBzK galaxies ($K_{{\rm AB}} < 23.2$ mag) 
and found that they reside in halos with a typical mass of 
$\sim3\times10^{11}\Msun$, which is comparable to the halo mass of 
the local disk galaxies (see also \citealt{ichikawa07}). 
This suggests that they are likely to be progenitors of the present-day 
disk galaxies if no major merge occurs at later time. 
With the SINS survey \citep{forster06}, \cite{forster09} studied 
the kinematical  properties of 80 galaxies 
at $1.3<z<2.6$ including BM/BX and sBzK galaxies. 
They derived velocity structure using H$\alpha$ emission line
and found that about one-third of their sample 
galaxies are rotation-dominated disks which follow a velocity-size 
relation similar to local disk galaxies. 
However, these galaxies show larger velocity dispersion than 
local disks, maybe suggesting the thicker disk. 
These observational results seem to suggest that star-forming 
galaxies at $z \sim 2$ are disk galaxies with a thicker disk. 
However, no study of the intrinsic shape of these 
galaxies has been established yet. 

The intrinsic structure of galaxies is related to the 
apparent axial ratio ($b/a$) or ellipticity ($1-b/a$) of galaxies. 
In the local universe, it is found that elliptical and disk galaxies have 
different observed axial ratio distributions. %\citep{lambas92, padilla08}. 
The distribution for local elliptical galaxies peaks at $b/a\sim0.8$ 
and decreases to zero at $b/a=0.2$, whereas disk galaxies 
show a rather flat distribution in the range of $b/a=0.2-0.8$ 
\citep[e.g.,][]{lambas92, padilla08, unterborn08}. 
In this paper, we constrain the intrinsic three-dimensional 
structure of star-forming BzK galaxies at $z\sim2$ from 
their observed axial ratios in order to investigate 
whether they indeed have a disk structure or not. 

This paper is organized as follows. 
Section \ref{sec:data} describes the optical-to-mid-infrared 
data sources used in this study. 
In Section \ref{sec:sampleselection}, we construct a sample of 
sBzK galaxies and derive their photometric redshifts 
and stellar masses. 
The morphological analysis is described in section \ref{sec:galfit}. 
Then we examine the intrinsic shape of the sBzK galaxies from their 
distributions of apparent axial ratios 
in section \ref{sec:intrinsicshape}. 
The possible origin and evolution of the intrinsic structure are 
discussed in section \ref{sec:discussion}. Conclusion is 
presented in section \ref{conclusion}. 
Throughout this paper, we use the AB magnitude system \citep{oke83} 
and assume a standard $\Lambda$CDM cosmology with parameters 
of $\Omega_m = 0.3$, $\Omega_{\Lambda}=0.7$, and $H_0 = 70$ km s$^{-1}$ Mpc$^{-1}$.

%----------------------------------------------------------------

\section{Data Sources}\label{sec:data}
We used the publicly available images in $U$, $B$, $V$, $R$, $I_c$,
 and \zmag~bands in the Great Observatories Origins Deep Survey North 
 (GOODS-N) field 
 \citep{capak04}\footnote{http://www.astro.caltech.edu/~capak/hdf/index.html}. 
The $U$-band image was obtained with the MOSAIC prime focus camera 
on the Kitt Peak National Observatory (KPNO) 4m telescope. 
The $B$-, $V$-, $R$-, $I_c$-, and \zmag-band images were obtained 
with Suprime-Cam attached to the Subaru telescope. 
Details of the data reduction can be found in \cite{capak04}. 
The seeing sizes of $U$-, $B$-, $V$-, $R$-, $I_c$-, and \zmag-band 
images are 1.\ar3, 1.\ar1, 1.\ar1, 1.\ar2, 1.\ar1, and 1.\ar1, 
respectively and the $5\s$ limiting magnitudes at $3\arcsec$ diameter 
aperture are 26.5, 26.0, 25.7, 25.8, 25.1, and 24.8 mag, 
respectively. 

In order to study the morphology of sBzK galaxies, 
we used the high resolution image 
observed with Advanced Camera for Surveys (ACS) on 
the Hubble Space Telescope (HST). 
The F850LP image (hereafter referred as \wz~image) 
was obtained from the HST/ACS v2.0 data products of the 
GOODS HST/ACS treasury program 
\citep{giavalisco04}\footnote{http://archive.stsci.edu/prepds/goods/} 
and covered an area of $\sim160$ arcmin$^2$. 
The pixel scale of the image is $0.\ar03$ pixel$^{-1}$. 
The full width at half maximum (FWHM) of the 
point-spread function (PSF) is $\sim0.\arcsec11$. 
The $5\s$ limiting magnitude of the \wz~image is 26.5 mag 
at 1.\ar0 diameter aperture. 

For the near-infrared (NIR) data, we used very deep and wide
 $J$-, $H$-, and $K$-band data obtained with Multi-Object 
InfraRed Camera and Spectrograph (MOIRCS; \citealt{suzuki08})
 on the Subaru telescope, i.e., MOIRCS Deep Survey 
(MODS; \citealt{kajisawa06, ichikawa07}). 
Four MOIRCS pointings cover $\sim70\%$ of the GOODS-N region 
($103.3$ arcmin$^2$, hereafter "wide field";
\citealt{kajisawa09}). 
One of the four pointings includes the Hubble Deep Field North 
(HDF-N; \citealt{williams96}) and is the deepest field of the MODS 
(hereafter deep field). 
The FWHM of the PSF is $0.\ar50$ for the deep field and 
$0.\ar61$ for the wide field. 
The images reach $J=25.3$, $H=24.6$, and $K=25.1$ mag 
($5\s$ at 1.\ar2 diameter aperture) for the wide field and 
$J=26.0$, $H=24.9$, and $K=25.7$ mag for the deep field. 

Mid-infrared images were obtained from deep observations 
with the Infrared Array Camera (IRAC) on the 
Spitzer Space Telescope (SST). 
We used 3.6 and 4.5\micron~band images from 
the First  (DR1) and Second Data Release (DR2) of the 
publicly available data provided by the SST Legacy Science 
program\footnote{http://ssc.spitzer.caltech.edu/spitzermission/observingprograms/legacy/goods/}. 
The pixel scale of all images after being drizzled is 0.\ar60
pixel$^{-1}$. 
The $5\s$ limiting magnitudes of the IRAC 3.6 and  
4.5\micron~images at $2.\arcsec4$ diameter aperture are 
25.4 and  25.3 mag, respectively. 

%--------------------------------------------------------------------------------------------------
\section{Sample Selection and SED fitting}\label{sec:sampleselection}

In order to select star-forming galaxies at $z \sim 2$ with BzK method, 
we first made position registrations of $B$- and $\zmag$-band 
images by using the $K$-band image as a reference. 
All images were homogenized to have seeing sizes of $1.\ar1$. 
Object detection was made in the $K$-band image with criteria 
of 5 connections above a $1.5\sigma$ minimum 
threshold by SExtractor version 2.5.0 \citep{bertin96}. 
For each object with {\tt MAG\_AUTO}$ < 24.0$ in $K$ band, 
the aperture photometry was made for $B$ and \zmag~bands at 
1.\ar6 diameter aperture by using dual-image mode of SExtractor, 
where the aperture size was found to provide the best signal-to-noise 
(S/N) ratio for the homogenized images. 
 
Star-forming galaxies at $z\sim2$ were selected by applying 
the BzK color criterion for star-forming galaxies (sBzK; \citealt{daddi04}) 
to the $K$-selected galaxies down to $K < 24.0$ mag. 
Since the response functions in our system are not identical to those 
used by \cite{daddi04}, we derived the color correction by convolving 
the empirical stellar spectra from \cite{pickles98} 
with the response functions including the detector's QE and atmospheric 
transmission. 
We found that $B-\zmag$ colors of stars in 
our system are bluer than those in \cite{daddi04} about $0.05-0.4$ mag 
depending on the spectral type, whereas our $\zmag-K$ colors are 
0.02 mag redder on average. 
In order to apply the BzK criterion consistent with \cite{daddi04}, we applied 
the $B-\zmag$ color correction to all objects by using the following equation;

\begin{equation}
\label {correctcolor}
(B-z)_{\rm Daddi} = (B-\zmag) + 0.4.
\end{equation}
Although the offset of 0.4 mag is not valid for all colors of $B-z$, it makes the 
selection criterion more secure as it is the upper limit offset. 
We did not correct for the $\zmag-K$ colors. 
The original sBzK criterion, $BzK \equiv (z - K) - (B-z)\geq -0.2$, was then 
applied to construct the sBzK sample. 
Objects with $z_{spec} < 1.4$ or $z_{spec} > 2.5$ are excluded 
from the sample based on spectroscopic identifications 
by \cite{wirth04} and \cite{barger08}. 
Resulting sample consists of 1029 sBzK galaxies. 

\begin{figure*}
\centering 
\begin{tabular}{cc}
\includegraphics[clip, angle=-90, width=7cm]{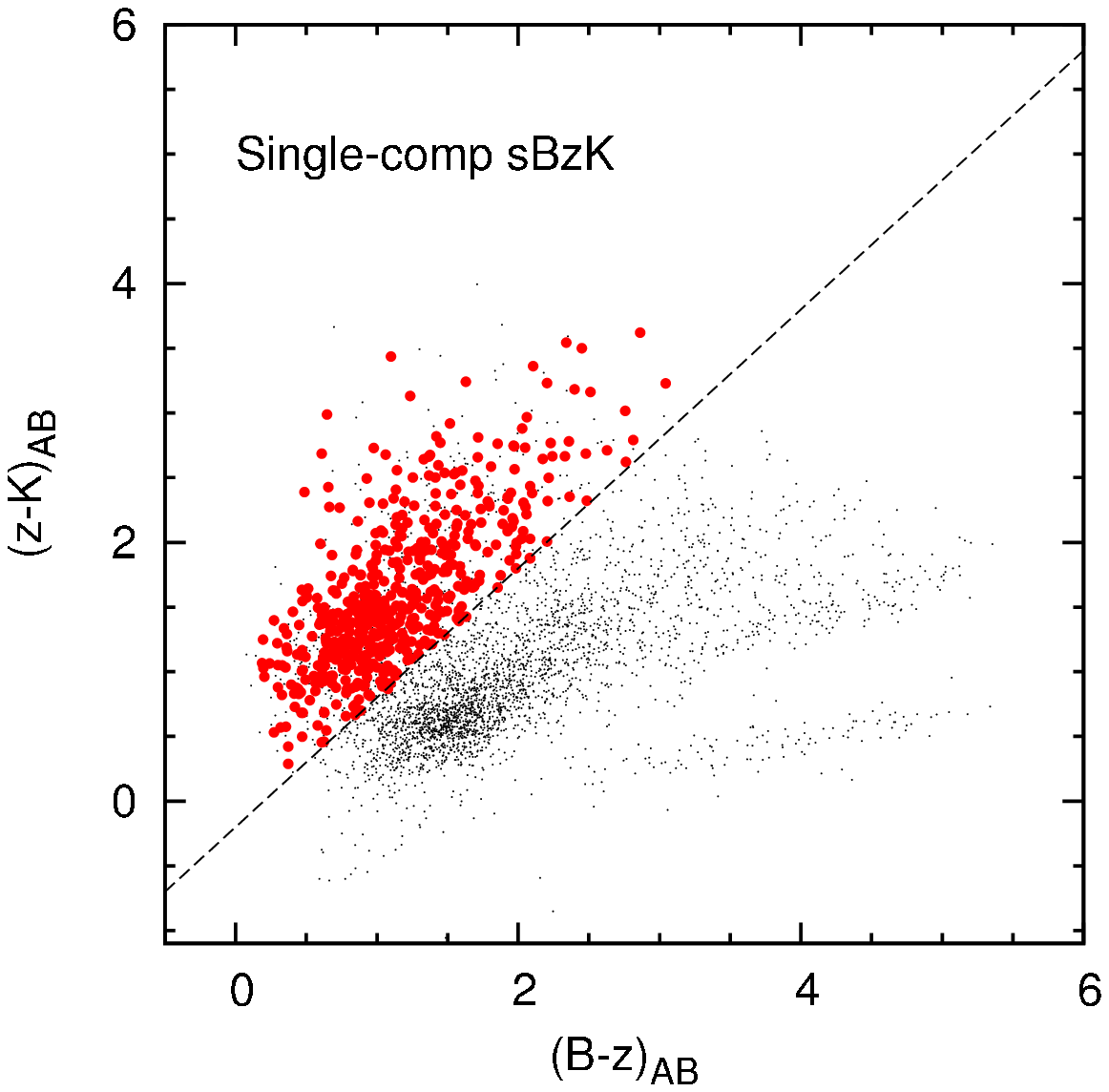}
\includegraphics[clip, angle=-90, width=7cm]{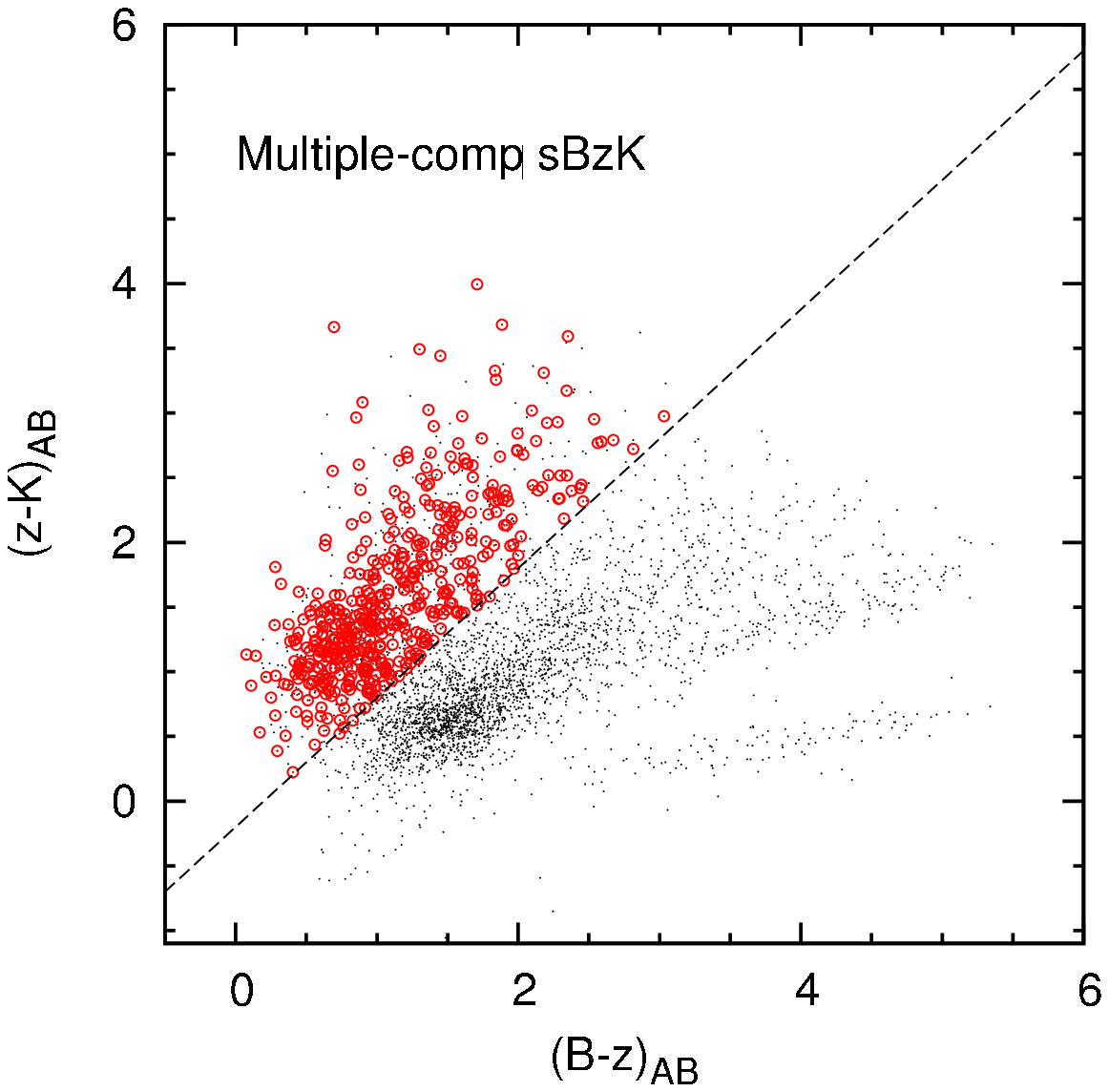}
\end{tabular}
\caption{
Two-color ($B-\zmag$) vs. ($\zmag - K$) diagram for sBzK galaxies in GOODS-N field. 
The $B-\zmag$ color shown here was corrected to match the filter system used by 
\cite{daddi04} (see text for more details). 
Red solid circles in the left panel represent 551 single-component sBzK galaxies, 
while red open circles in the right panel show 478 multiple-component 
objects (section \ref{sec:sampleselection}). 
All objects with $K<24.0$ mag are also shown with dots. 
}
\label{bzk}
\end{figure*}

\begin{figure*}[ht]
\centering 
\begin{tabular}{c c}
\includegraphics[clip, angle=-90, width=7cm]{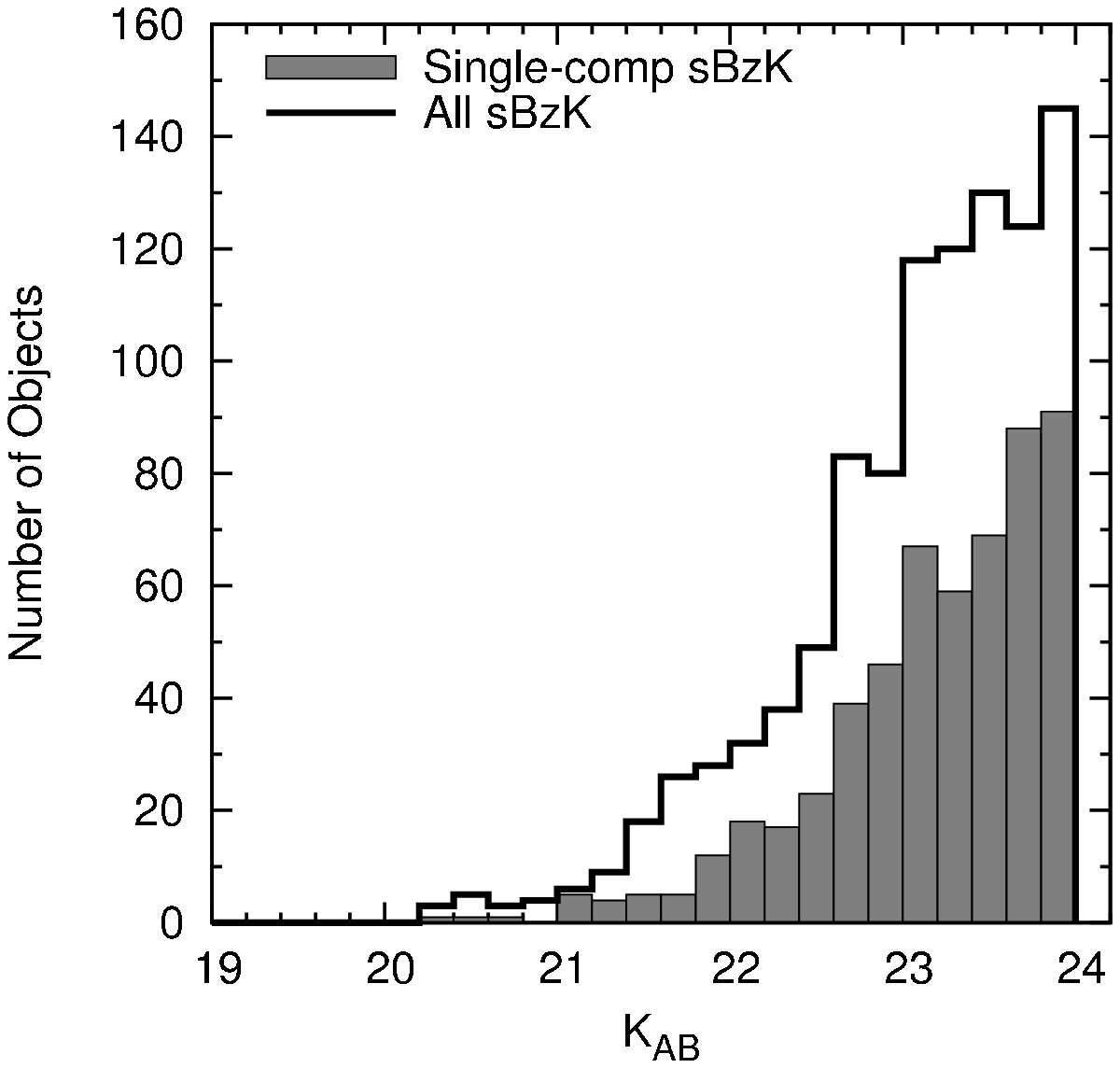}
\includegraphics[clip, angle=-90, width=7cm]{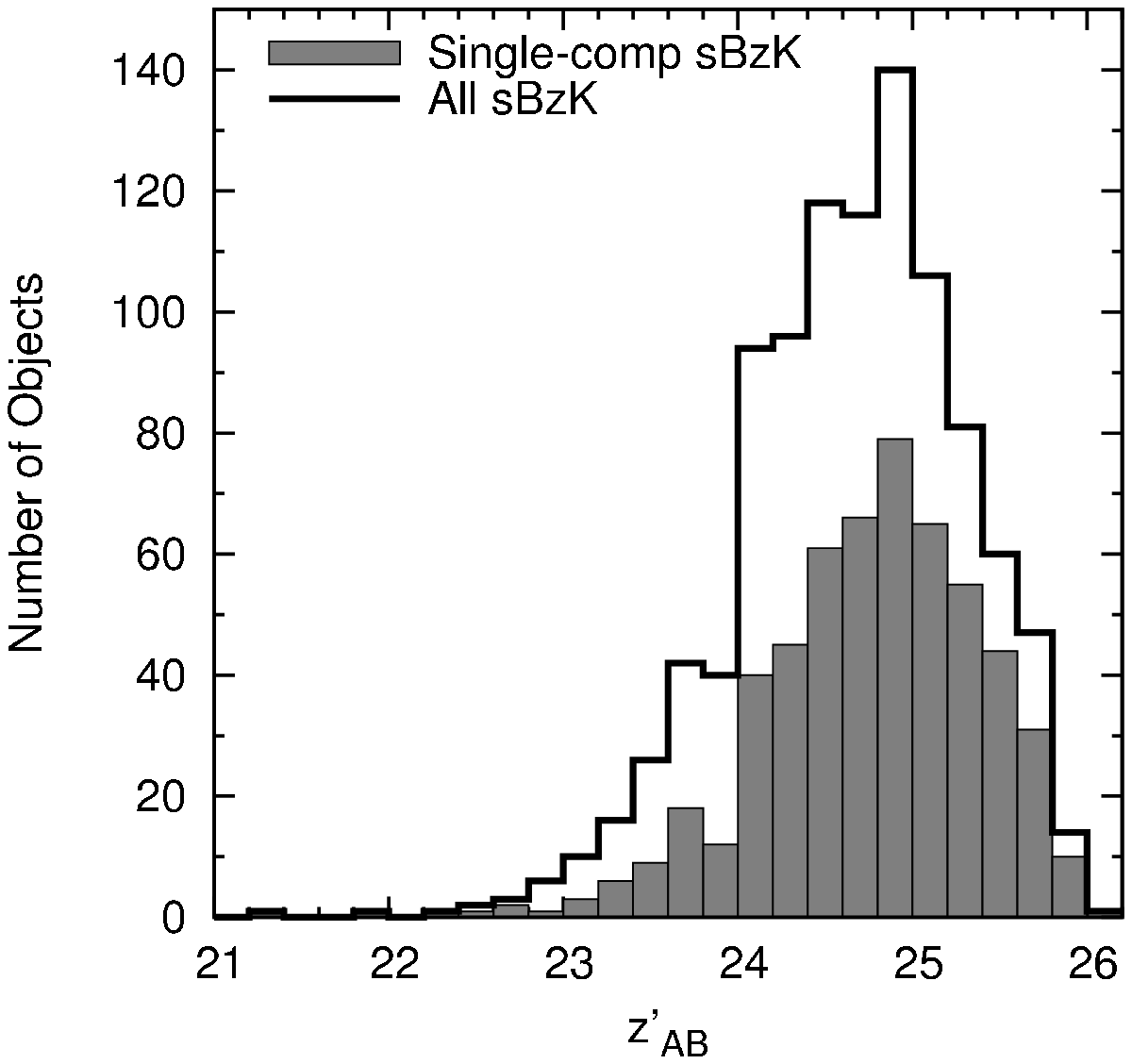}
\end{tabular}
\caption{Magnitude histograms of all and single-component sBzK galaxies in $K$ band 
(left panel) and \zmag~band (right panel). Open histograms are for 
all sBzK galaxies after removing objects with $z_{spec} < 1.4$ or $z_{spec} > 2.5$, 
while black shaded histograms represent the single-component sBzK galaxies. 
}
\label{mag}
\end{figure*}

In order to study the morphological properties, 
we cross-matched the sBzK galaxies with the high-resolution \wz~image, 
for which the central wavelength corresponds to the 
rest-frame UV wavelength of $\sim3000$\AA~at $z\sim2$. 
Counterparts of the sBzK galaxies in the \wz~images are identified 
if they are detected above $3\sigma$ limiting magnitude (27 mag) 
within 0.\ar72 radius centering at the positions of sBzK galaxies 
in the $K$-band image. This radius is twice as large as 
the median $1/2$FHWM of the sample. 
The sBzK galaxies are divided into 2 groups: 
sBzK galaxies matched with only one object in the \wz~image 
(hereafter single-component objects) and sBzK galaxies matched 
with more than one object (hereafter multiple-component objects). 
551 galaxies are classified as single-component objects and 
478 galaxies are multiple-component objects; 54\% of the sBzK galaxies 
show a single structure in the \wz~image. 
Figure \ref{bzk} indicates that the single-component and the 
multiple-component sBzK galaxies show very similar distributions 
in the BzK diagram (left and right panel, respectively). 
In this paper, we concentrate our analysis on single-component objects 
(red circles in the left panel of Figure \ref{bzk}). 
$K$ magnitudes and \zmag~magnitudes of the single-component galaxies, 
which are shown in Figure \ref{mag}, distribute in the range 
of $20.0-24.0$ mag and $21.0-26.0$ mag, respectively. 
The distributions of all sBzK galaxies are 
also shown in both panels with open histograms. 
It is worth noting that as star-forming regions in a face-on galaxy 
could be seen separately in the rest-frame UV wavelength, 
it is possible that they are selected as a multiple-component object in this analysis. 
Study of structure at the rest-frame optical wavelength 
in the future is helpful to see if they really have multiple components or 
just star-forming regions in a face-on galaxy. 

\begin{figure*}
\centering
\begin{tabular}{c c}
\includegraphics[clip, angle=-90, width=7cm]{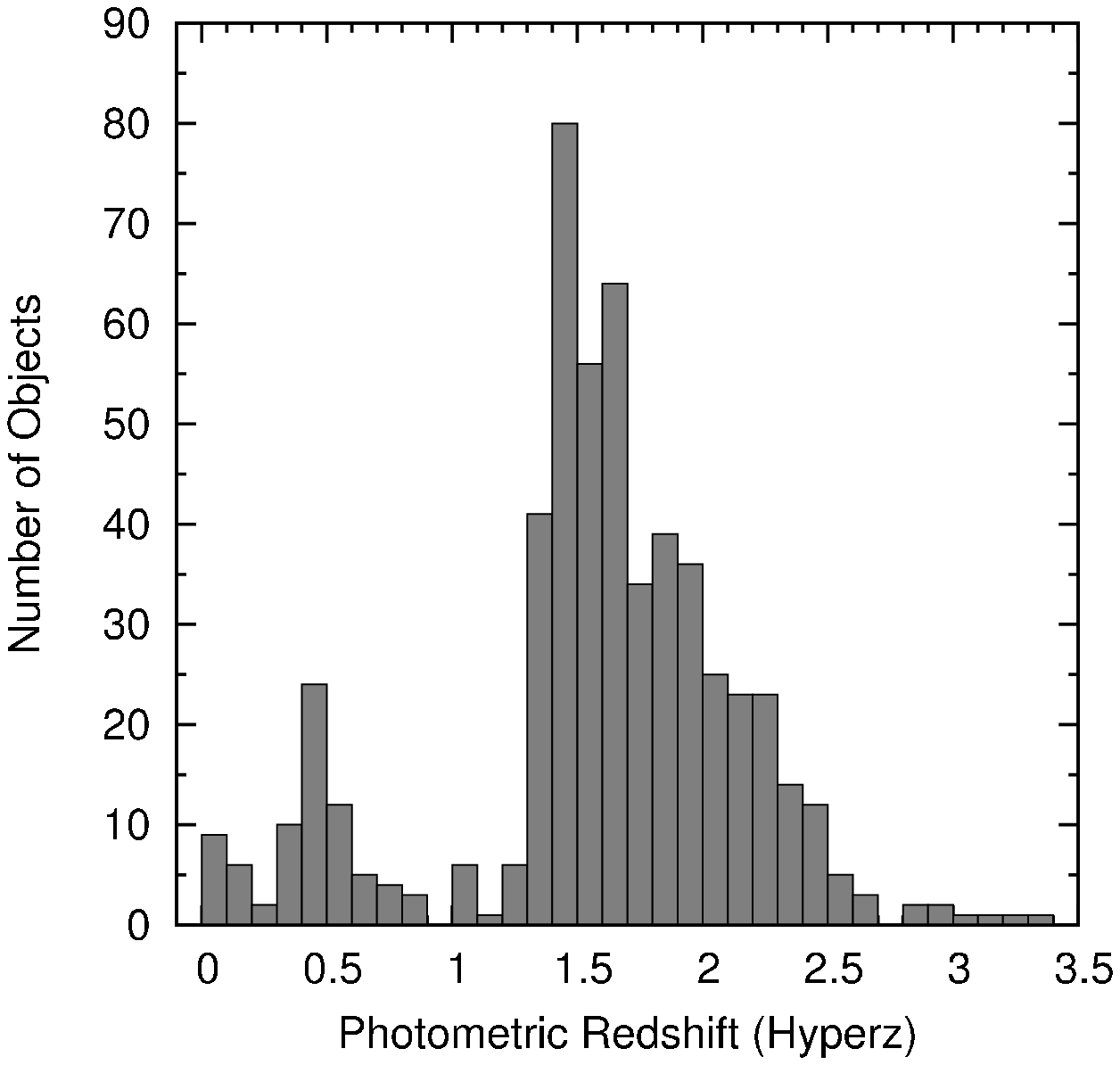}
\includegraphics[clip, angle=-90, width=7cm]{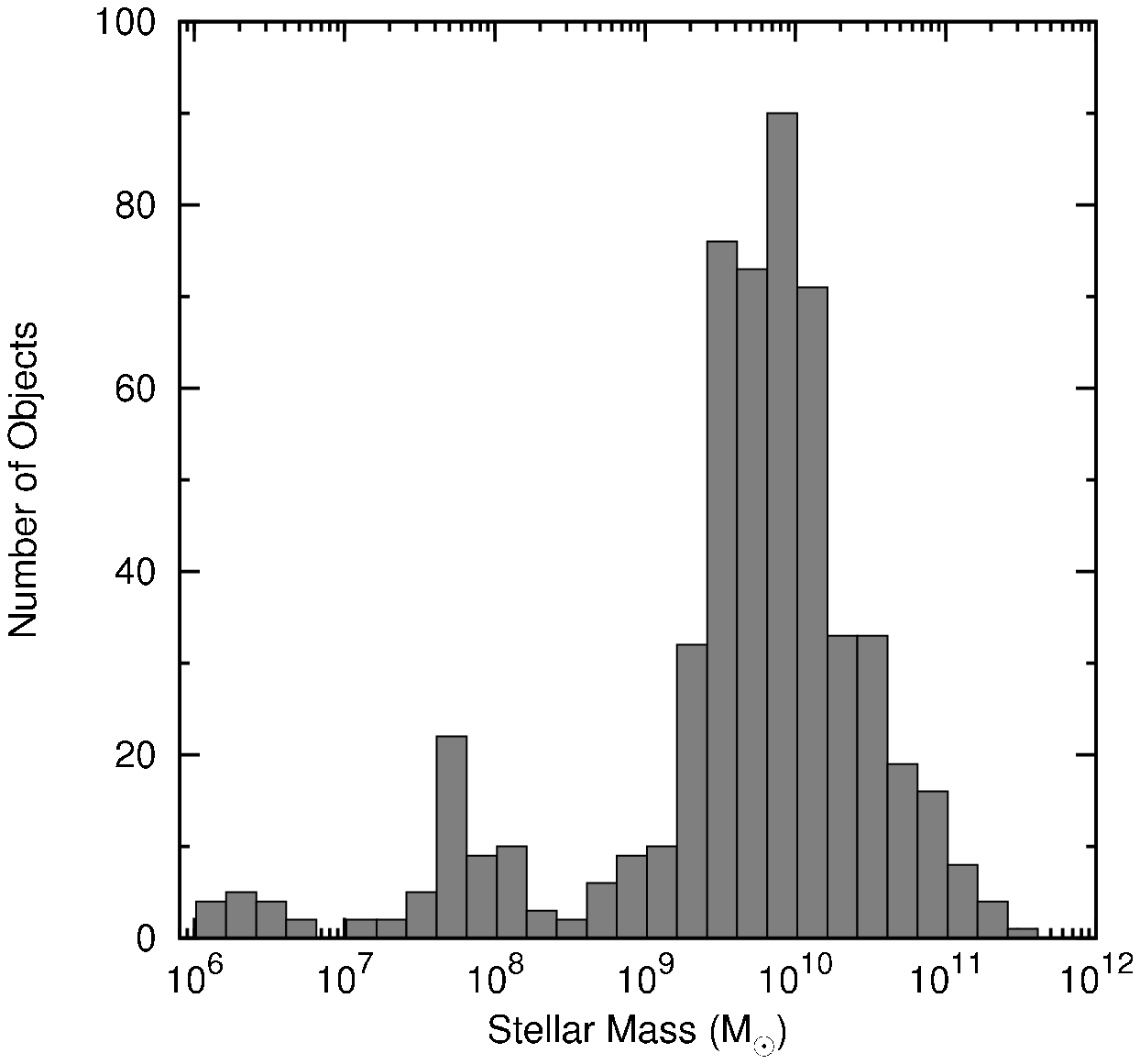}
\end{tabular}
\caption{
Histograms of photometric redshifts (left panel) and stellar masses (right panel) 
of the single-component sBzK galaxies. 
}
\label{photomass}
\end{figure*}

The photometric redshifts and stellar masses of the single-component 
sBzK galaxies were obtained by spectral energy distribution (SED) fitting. 
SEDs of the single-component sBzK galaxies were 
constructed by the photometry in $U$, $B$, $V$, $R$, $I_c$, \zmag, $J$, $H$, 
$K$, IRAC 3.6\micron, and 4.5\micron~bands. For all but IRAC images, 
we smoothed the images to match the seeing size of $U$-band images 
($1.\ar3$) as it has the worst seeing size. Then the aperture photometry 
was made at $1.\ar6$ diameter aperture by using dual-image mode.  
The total magnitudes were obtained by using the correction 
factor that was determined by scaling the aperture magnitude of 
each object to the {\tt MAG\_AUTO} in the $K$-band image. 
For the IRAC images, we made the aperture photometry without 
homogenizing the images at 2.\ar4 diameter aperture, which maximizes the S/N. 
The total magnitudes were obtained by 
using the aperture correction factor determined for each band.
The process of determining the aperture correction factor is identical 
to that used by \cite{yuma10}. The aperture photometry at the same 
aperture size as done for our targets was made for artificial objects with known total 
magnitudes put into the IRAC images. The correction 
factors are $-0.62$ mag and $-0.72$ mag for 3.6\micron~and 4.5\micron~
bands, respectively. 
The photometric redshifts 
of the single-component sBzK galaxies were obtained with $Hyperz$ 
\citep{bolzonella00}. 
We checked the accuracy of our photometric redshift estimation 
using objects with available spectroscopic redshifts \citep{barger08}. 
The comparison between the photometric and the spectroscopic 
redshifts shows median and standard deviation of 
$\delta_z\equiv(z_{phot}-z_{spec})/(1+z_{spec})$ of $-0.03$ and 0.07, respectively. 
The histogram of the resulting photometric redshifts is shown in the left panel 
of Figure \ref{photomass}. 
Most of the samples have the photometric redshifts between 1.3 and 2.5. 
Note that although we did not exclude sBzK galaxies with 
$z_{phot} < 1.3$ or $z_{phot} > 2.5$ from our sample, 
it does not affect the results. 
The stellar mass was obtained from SED fitting with $z_{phot}$. 
We constructed model SEDs of various star-formation histories 
(i.e., instantaneous burst, constant star formation, or 
exponentially declining models) by using the \cite{bc03} synthesis
code. \cite{salpeter55} initial mass function (IMF) with a mass range 
of $0.1-100$ \Msun~was assumed. 
The metallicity was fixed at solar abundance. 
The dust attenuation law of \cite{calzetti00} was adopted 
and $E(B-V)$ range was taken from 0.0 to 1.0 mag with a step of 0.01 mag. 
The SED fitting process was the same as that used by \cite{yuma10}. 
The histogram of stellar masses is shown in the right panel of Figure \ref{photomass}. 
It is seen that most of the single-component sBzK galaxies have the 
stellar masses of $10^9-10^{11}$ \Msun. 

%----------------------------------------------------------------------------------------------%
\section{Morphological Analysis}\label{sec:galfit}
\subsection{Estimation and Accuracy of S\'ersic parameters}
We determined the structural parameters of the single-component 
sBzK galaxies by fitting the two-dimensional light distributions 
in the \wz~image with a single S\'ersic profile
\citep{sersic63, sersic68}. 
The S\'ersic profile is expressed by 
$I(r) = I_e \exp
\left\{-\kappa_n\left[\left(\frac{r}{r_e}\right)^{1/n}-1\right]\right\}$, 
where $I_e$ is the surface brightness at the effective radius $r_e$ 
and $\kappa_n$ is a parameter related to the S\'ersic index $n$. 
The index $n$ determines the shape of the profile. 
In the local universe, an exponential profile ($n=1$) is seen 
for a disk component, whereas $n\sim4$ is representative of 
a spheroid component. 
The two-dimensional profile fitting was done by using the GALFIT 
version 2 \citep{peng02}. 
There are seven free parameters which are determined by $\chi^2$ 
minimization: central position, total magnitude, $n$, $r_e$, 
apparent axial ratio ($b/a$), and position angle. 
The S\'ersic index $n$ was initially set to be $n=1.5$. We tested the 
dependence of final GALFIT results on the initial S\'ersic index by 
varying the initial $n$ from 1.0 to 4.0 and found that varying 
the initial S\'ersic index does not affect the final fitting results ($\Delta n/n < 0.05$). 
Other initial guess parameters for the fit are outputs from the SExtractor. 
The noise for each pixel required for deriving the errors is 
determined from the variance maps produced during the drizzling process. 
The PSF used to be convolved with the S\'ersic model was obtained 
by stacking images of unsaturated stars with stellarity 
index larger than 0.98 and without nearby objects in the \wz~image. 

\begin{figure*}[ht]
\centering
\includegraphics[clip, width=16cm]{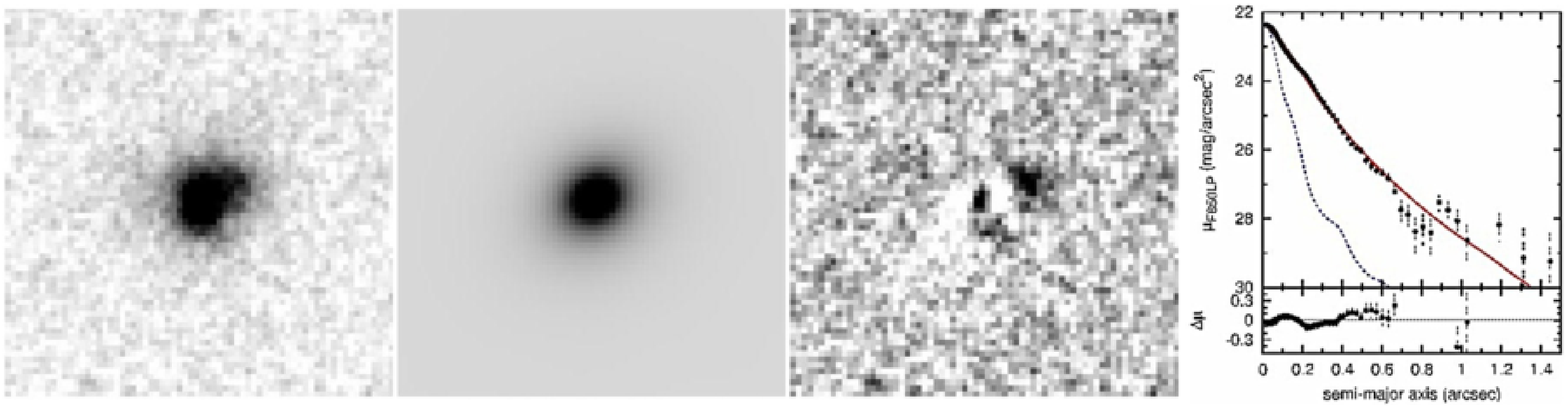}\\
\includegraphics[clip, width=16cm]{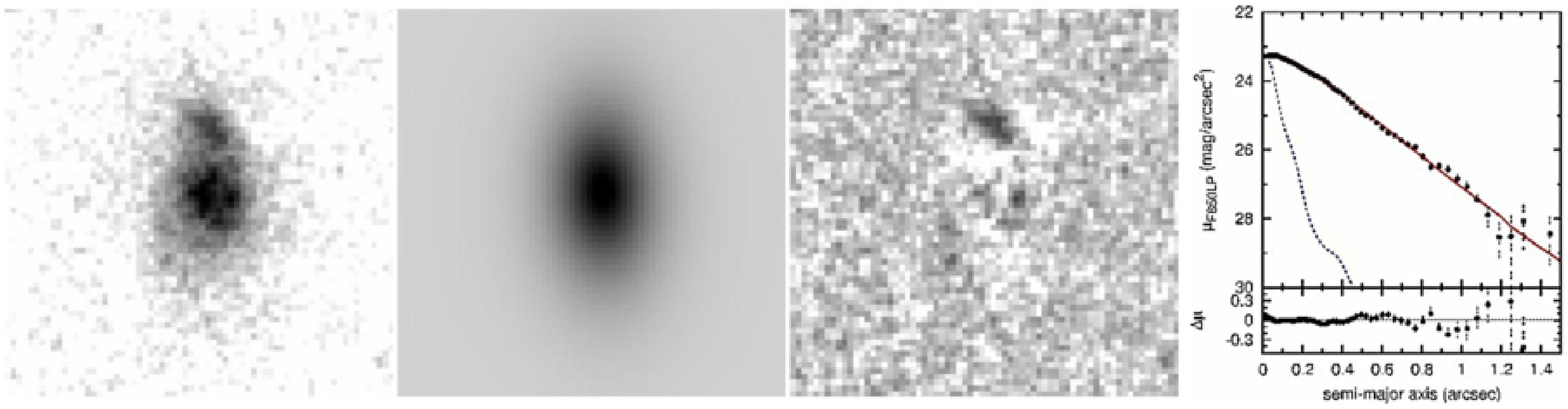}\\
\includegraphics[clip, width=16cm]{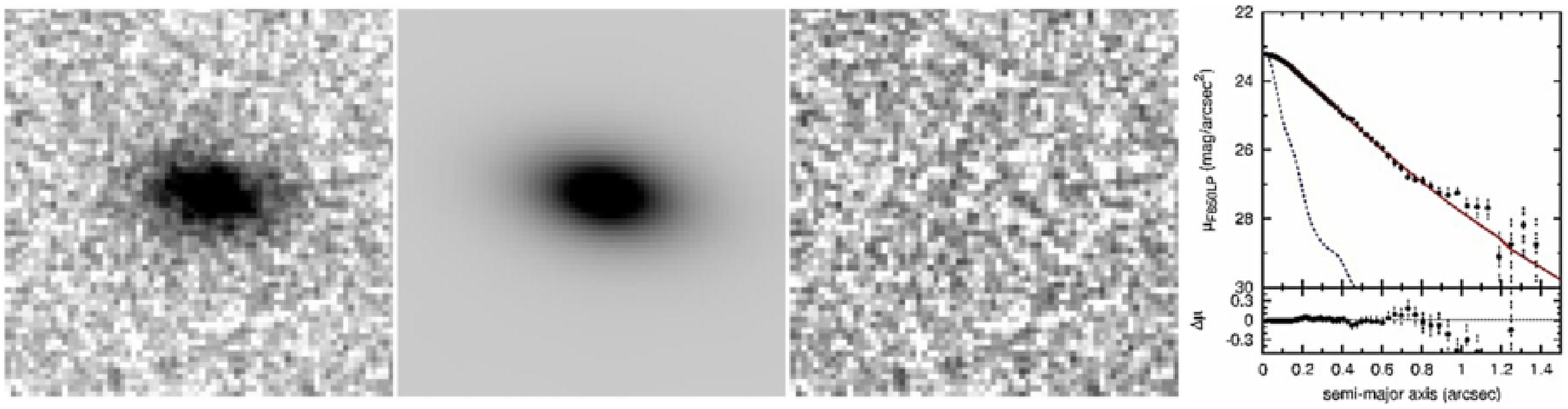}\\
\includegraphics[clip, width=16cm]{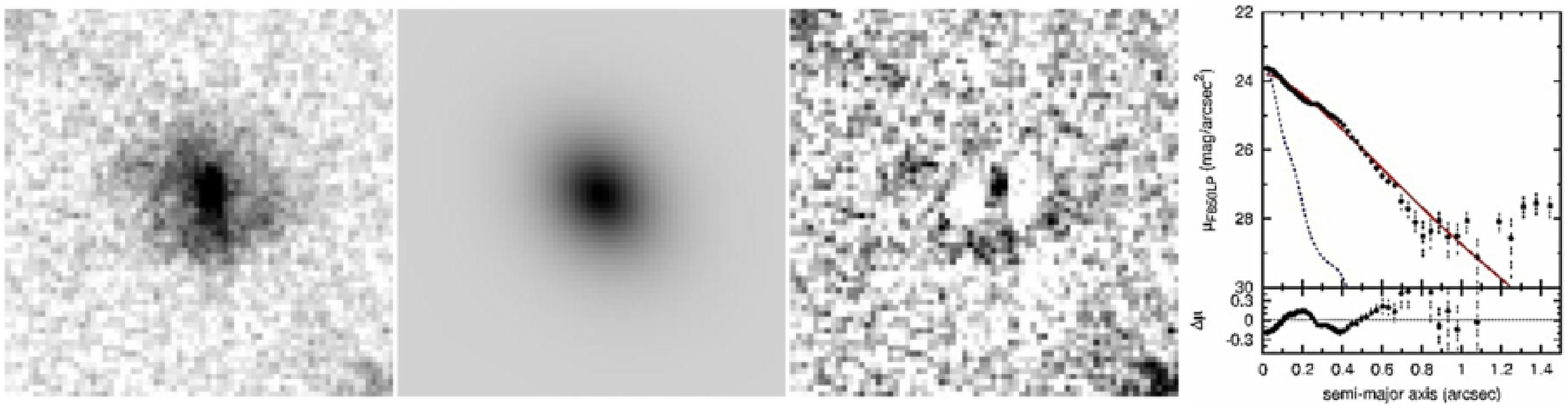}\\
\caption{Examples of the 2D surface brightness modeling with GALFIT for the 
single-component sBzK galaxies. The first (from the left), second, and third panels 
show the ACS/F850LP image of sBzK galaxies, the best-fitting model constructed 
by GALFIT, and the residual image, respectively. North is at the top and east is to the left. 
The size of each image is $2.\ar0\times2.\ar0$ corresponding to$\sim17\times17$ 
kpc at $z=2$. To clearly display the residual part, the residual images are shown 
with narrower scale range. The rightmost panel shows the azimuthally averaged 
1D surface brightness profile of each object (black circles). 
Red solid and blue dot lines indicate the best-fitting model profile and 
the PSF, respectively. The bottom of the rightmost panel represents the 
residual profile of  surface brightness.
}
\label{galresult}
\end{figure*}

Examples of GALFIT results are shown in Figure \ref{galresult}. 
We excluded objects with incredibly large errors obtained from GALFIT, 
which comprise about 9\% of the single-component sBzK galaxies. 
The unreasonably large errors of some objects indicate the divergence 
in the fitting process. 
The number of the remaining single-component sBzK galaxies is 499. 

We carried out the Monte Carlo simulations in order to examine 
the accuracy and reliability of the parameters derived by GALFIT. 
About 5000 artificial objects were generated for disks ($n=1$) 
and spheroids ($n=4$), with uniformly distributed random 
magnitudes ($21 - 26$ mag), 
random effective radii ($0.\ar01-1.\ar00$), ellipticities 
($0.1-1.0$), and position angles ($0-180^{\circ}$). 
The magnitude range was adopted so as to be the same as 
that of our sample. 
The artificial galaxies were then convolved with PSF of the
\wz~image and inserted into the observed \wz~image. 
GALFIT was performed to the artificial objects in the manner identical
 to that used for the sBzK galaxies. 

\begin{figure*}
\centering
\begin{tabular}{c c}
\includegraphics[clip, angle=-90, width=7cm]{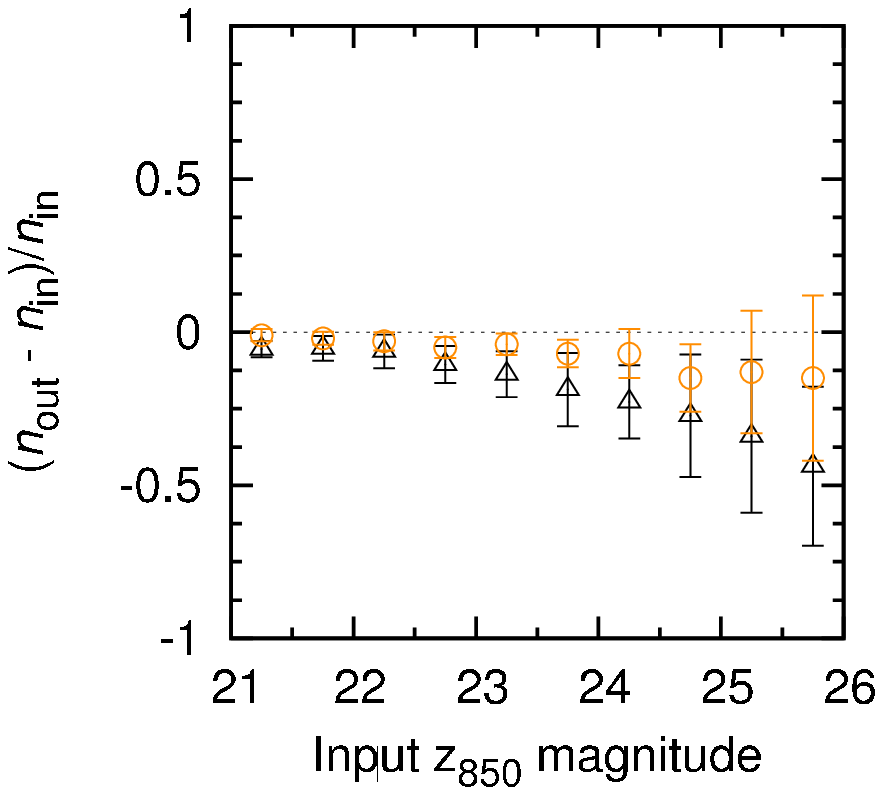}&
\includegraphics[clip, angle=-90, width=7cm]{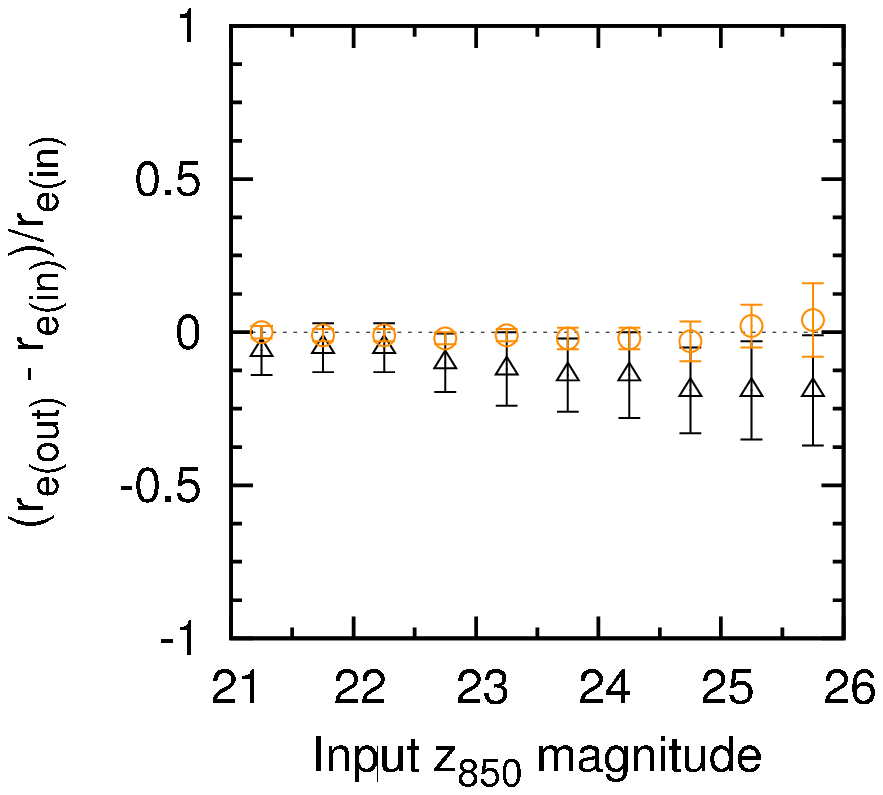}\\
\end{tabular}
\caption{Accuracy of the output parameters from GALFIT based on 
the Monte Carlo simulations as a function of magnitude for 
the recovered median S\'ersic index (left panel) and median 
effective radius (right panel). Open circles and triangles refer 
to artificial objects with an exponential profile ($n=1$) and those 
with a spheroid profile ($n=4$), respectively. Error bars correspond 
to 1$\sigma$ of the distribution.}
\label{sim_mag}
\end{figure*}

Figure \ref{sim_mag} shows the resultant accuracy of the output parameters 
from Monte Carlo simulations with $1\sigma$ of distributions as a function of 
the input \wz~ magnitude. 
The left panel represents the accuracy of the recovered S\'ersic indices. 
Over all range of magnitudes in the simulations ($21-26$ mag), 
the output $n$ distribution has a median value of $\langle n\rangle = 0.96$
 with $\sigma_n=0.68$ for the disks and $\langle n\rangle =3.49$ 
with $\sigma_n=1.34$ for the spheroids. 
Typical errors on $n$ obtained from GALFIT are 0.02 and 0.06 
for the simulated disks and spheroids, respectively. 
The recovered $n$ is significantly underestimated for the 
objects with $n=4$; however, it is possible to distinguish two populations 
by classifying objects with $n\geq2.5$ as spheroid-like and 
those with $n<2.5$ as disk-like down to $\wz=24.5$ mag. 
At $\wz \sim 25.5$ mag, about half of the input objects with $n=4$ 
show output $n \leq 2.5$, but a fraction of the single-component 
sBzK galaxies fainter than 25.5 mag is very much small. 
Nevertheless, keeping this in mind, we also examine 
the brighter subsample in the subsequent analysis (section \ref{subsec:intrinsic}). 
The right panel of Figure \ref{sim_mag}, which shows 
the accuracy of the recovered $r_e$, 
indicates that the disk profiles are better recovered than 
the spheroid profiles. 
The effective radii of disk profiles are well recovered down to 
magnitude of 26.0 mag. 
The typical errors in determining the effective radius are 3\% for 
the disks and 12\% for the spheroids in the magnitude range of $20.0-26.0$. 

%---------------------------------------------------------------------------------------
\subsection{Results on S\'ersic Parameters}
\label{subsec:nre}

\begin{figure*}
\centering 
\begin{tabular}{cc}
\includegraphics[clip, angle = -90, width = 7cm]{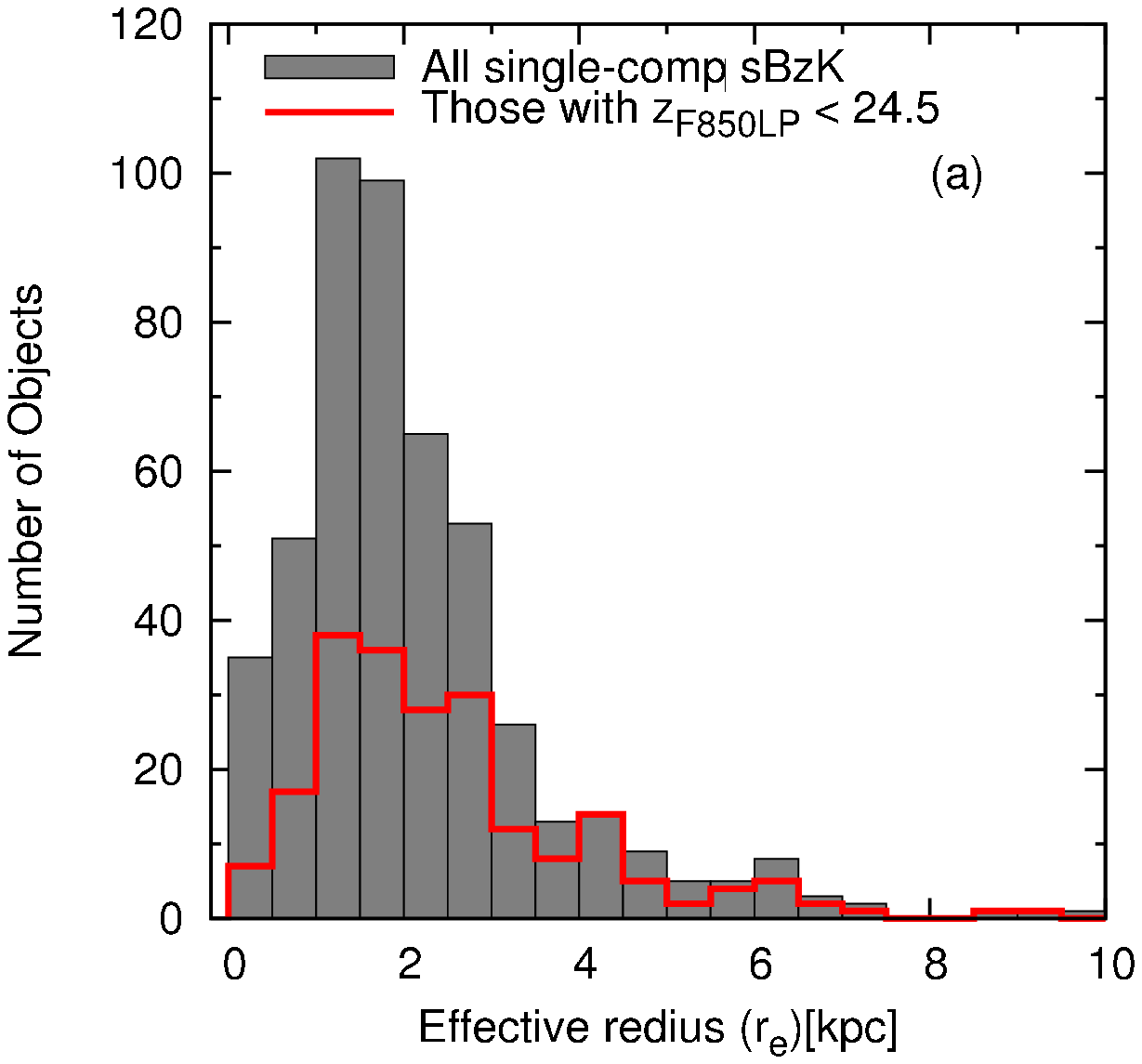} & 
\includegraphics[clip, angle = -90, width = 7cm]{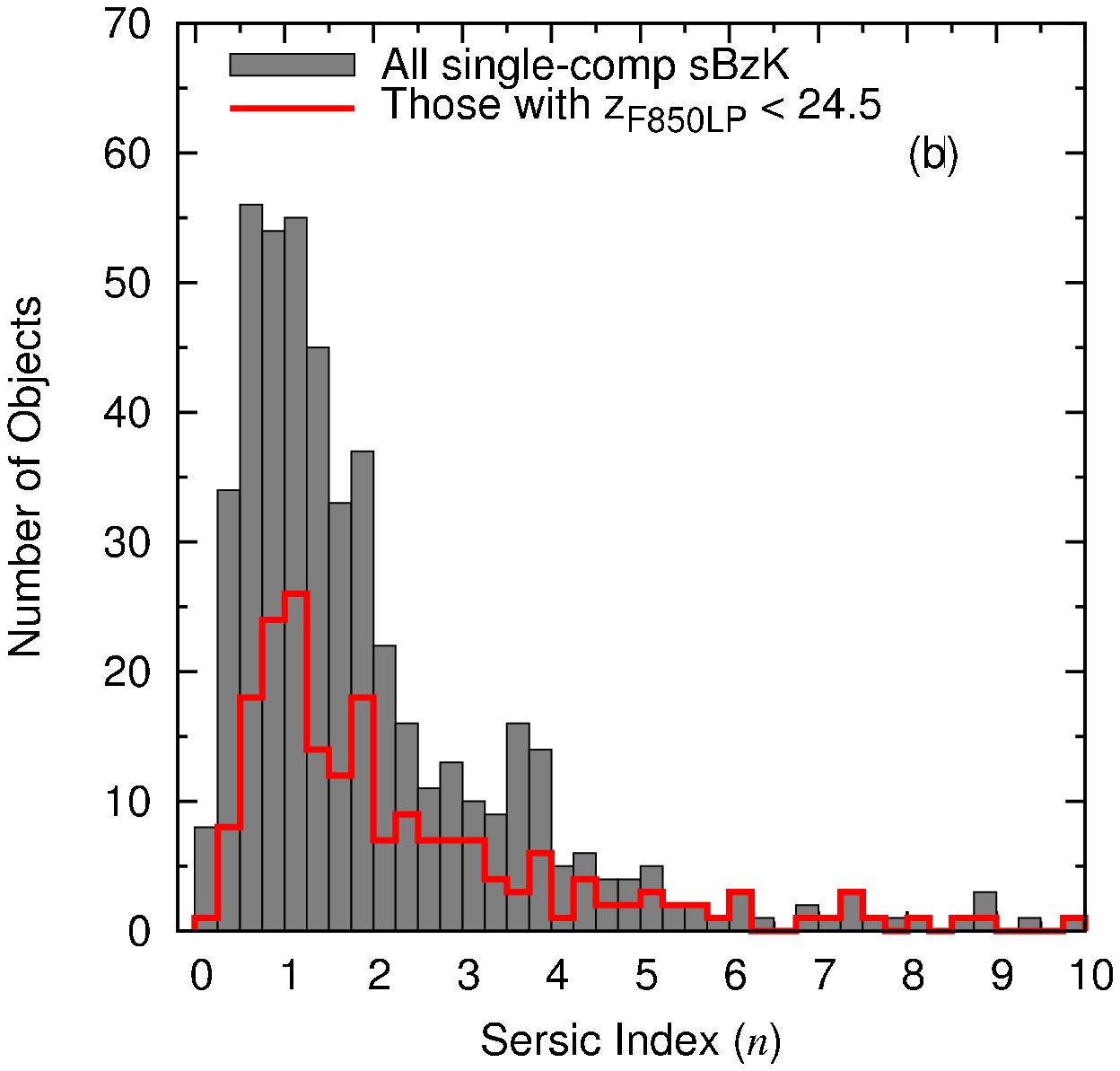} \\
\end{tabular}
\caption{Distributions of effective radius $r_e$ (left panel) and S\'ersic index $n$ (right panel). 
All 499 single-component sBzK galaxies are shown in the solid black histogram, 
whereas those with $\wz <24.5$ mag are in the open red histogram. }
\label{dist}
\end{figure*}

Distributions of effective radius ($r_e$) and S\'ersic index ($n$) are
 shown in Figure \ref{dist}. 
The effective radius is described in kpc calculated with 
the photometric redshift. 
Most of the single-component sBzK galaxies show the effective radius 
around $1-3$ kpc (Figure \ref{dist}(a)), which is 
consistent with the previous studies of star-forming 
galaxies at $z=2-4$ \citep{overzier10, swinbank10}. 
We also show in the figure the distribution for the galaxies with 
$\wz<24.5$ mag, which is the limiting magnitude at which the 
structural parameters are well recovered. It is seen that 
the distributions of both all single-component sBzK galaxies and 
those with $\wz<24.5$ mag are very similar. 
In Figure \ref{dist}(b), the distribution of derived S\'ersic index 
peaks at $n\sim1$ for both samples. If we refer to a galaxy with $n\geq2.5$ as 
a "spheroid-like" object and that with $0.5 \leq n < 2.5$ as a "disk-like" object, 
we find that 64\% (318/499) of the single-component sBzK galaxies 
are disk-like objects. 

\begin{figure}
\centering
\includegraphics[clip, angle=-90, width=7cm]{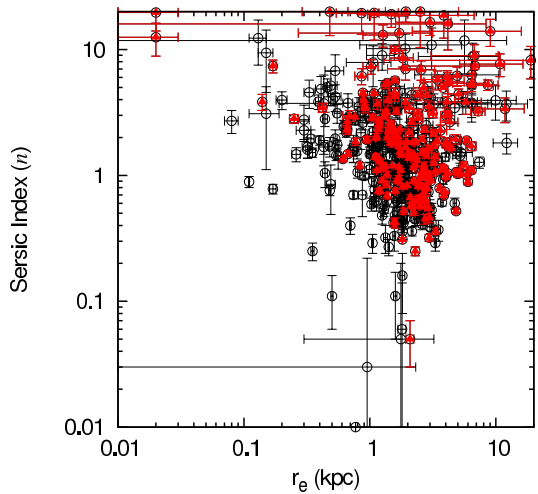}
\caption{S\'ersic index $n$ versus effective radius $r_e$. 
Open circles represent all 499 single-component sBzK galaxies and 
solid triangles show those with $\wz<24.5$ mag. }
\label{nre}
\end{figure}

In Figure \ref{nre}, we plot S\'ersic index $n$ against effective radius $r_e$. 
A slight anti-correlation between the effective radii and 
the S\'ersic indices is seen; 
galaxies with larger size show smaller S\'ersic index. 
Such a trend can also be seen among Lyman alpha emitters (LAEs) 
at $z=3.1$ in the rest-frame UV wavelength \citep{gronwall10}. 
It is interesting to note that the trend is different from the 
correlation for early-type galaxies 
in the local universe, where the S\'ersic index increases with
increasing size \citep{caon93, donofrio01, graham03, 
aguerri04, rothberg04, aceves06, kormendy09, ascaso11}.

\begin{figure}
\centering
\includegraphics[clip, width=7cm]{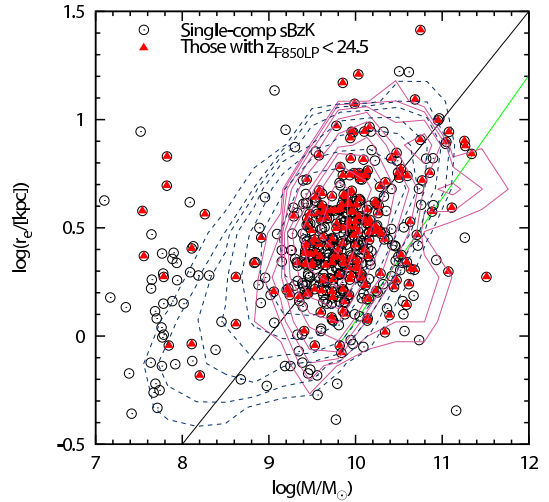}
\caption{
Effective radius and stellar mass of the single-component sBzK galaxies 
with $0.5 \leq n < 2.5$. 
The effective radius was corrected according to the ratio of the median 
effective radius in the rest-frame optical to that in the rest-frame UV 
wavelengths (see text). 
The stellar masses of all samples are corrected to the Salpeter IMF. 
Dashed and solid contours indicate the distributions of $z\sim0$ and 
$z\sim1$ disk galaxies by \cite{barden05}, respectively. 
The black solid line represents the average surface density measured from 
the disk galaxies at $z=0-1$ (log$\Sigma_M = 8.50$ with $q=0.5$; \citealt{barden05}). 
Green line shows the size-mass relation of early-type galaxies from 
the Sloan Digital Sky Survey (SDSS) by \cite{shen03}. 
}
\label{surfaceden}
\end{figure}

Figure \ref{surfaceden} shows the effective radius against stellar 
mass of the single-component sBzK galaxies with $0.5 \leq n <2.5$ 
and the relations for local galaxies. 
Since our analysis is based on the \wz~image which corresponds to 
the rest-frame UV wavelength ($\sim3000$ \AA) at $z\sim2$, 
it is difficult to compare the results for sBzK galaxies with those for the local 
galaxies which have been studied in the optical wavelength. 
In order to make more useful comparison, we corrected the 
effective radius in the rest-frame UV to that in the rest-frame optical 
wavelength with $r_{e,opt}/r_{e,UV} =1.37$, 
the ratio of median effective radius in the rest-frame UV and 
optical wavelengths for star-forming galaxies 
(BM/BX and Lyman break galaxies) 
at the similar redshift by \cite{swinbank10}. 
The stellar masses of the local galaxies are corrected to the 
Salpeter IMF. 
It is seen from the figure that the corrected effective radius of 
the $z\sim2$ sBzK galaxies slightly depends on their 
stellar mass; size increases with the stellar mass. 
Most of the single-component sBzK galaxies distribute 
in the region where the local disks or $z\sim1$ disks studied 
by \cite{barden05} locate (contours). 
In other words, most of the sBzK galaxies show the surface stellar mass 
density comparable to the local disks, though some of them 
show larger surface stellar mass densities that are more 
typical to the local elliptical galaxies (green line; \citealt{shen03}). 
These results seem to suggest that most of the single-component 
sBzK galaxies at $z\sim2$ already have a disk structure close to 
that of the present-day disk galaxies. 

%------------------------------------------------------------------------------------------Intrinsic shape
\section{Distribution of Apparent Axial Ratio ($b/a$) and Intrinsic Shape of the sBzK Galaxies}
\label{sec:intrinsicshape}

\subsection{$b/a$ Distribution of the Single-Component sBzK Galaxies}
\label{subsec:qdist_bias}

The peak of S\'ersic-index distribution at $n\sim1$ suggests 
that most of the single-component sBzK galaxies are disk-like. 
The nature of $n \sim 1$ is usually interpreted as an evidence for 
a disk galaxy.
However, it is important to investigate whether they indeed 
have a disk shape or not. 
In the local universe, the intrinsic three-dimensional shape is 
examined statistically from the distributions of the apparent 
axial ratios $b/a$ \citep[e.g.,][]{ryden04, vincent05, padilla08, unterborn08}. 
In this study, we examine the distribution of the apparent
axial ratios of the single-component sBzK galaxies, and
compare it with model distribution to constrain their intrinsic 
shape. 

\begin{figure}
\centering 
%\begin{tabular}{cc}
\includegraphics[clip, angle = -90, width = 7cm]{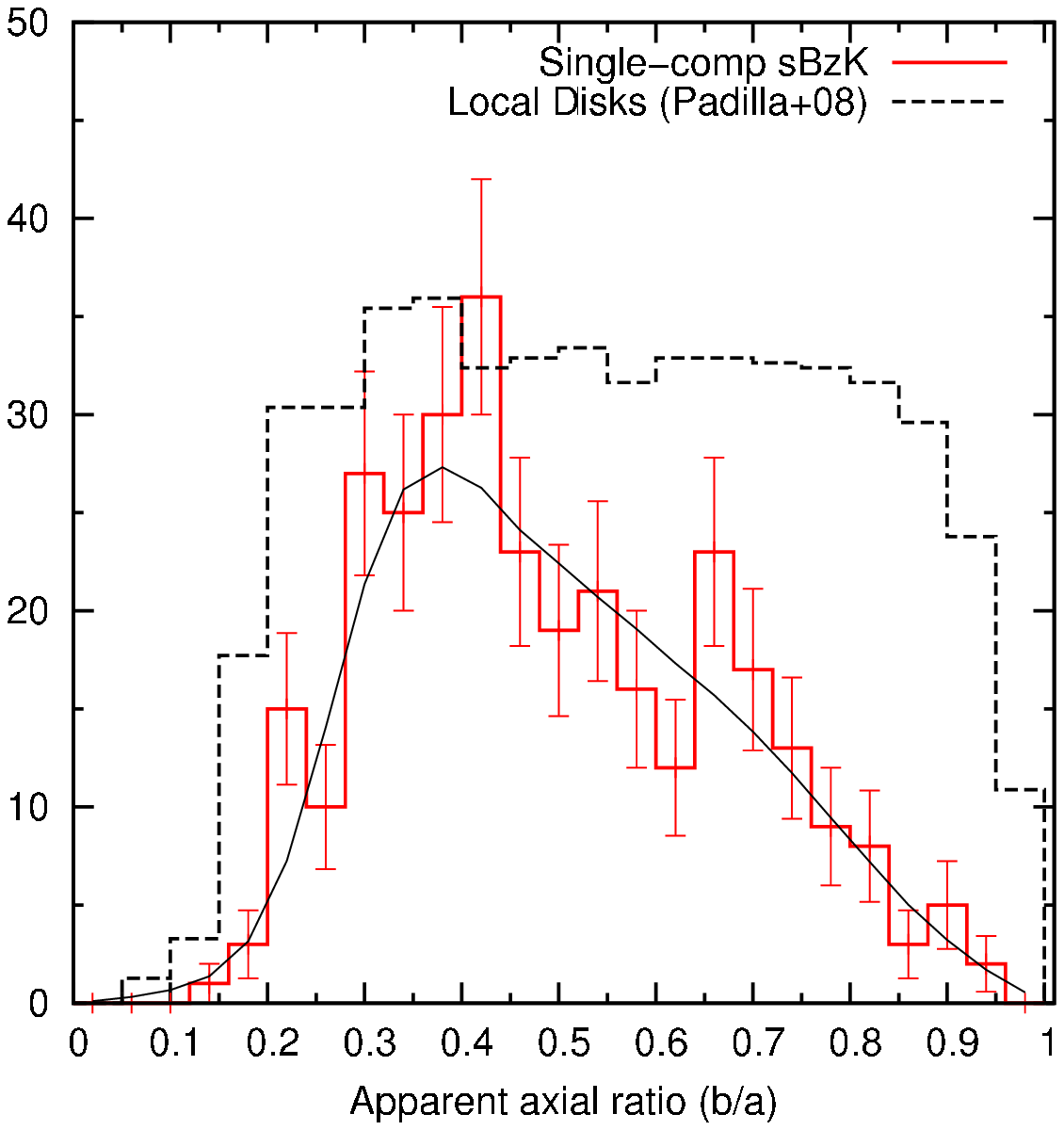}
\caption{Distributions of the apparent axial ratios ($b/a$) of the 
single-component sBzK galaxies and the local disk galaxies \citep{padilla08}. 
The observed $b/a$ distribution of sBzK galaxies with $0.5 \leq n < 2.5$ 
is shown with red solid histogram and the 
best-fitting model distribution is indicated by black solid line. 
Dashed line represents the observed $b/a$ distribution of the local disk 
galaxies \citep{padilla08}. 
The distributions are normalized so that they have the same peak. 
} 
\label{alldistq}
\end{figure}

The $b/a$ distribution of the single-component sBzK galaxies 
with $0.5 \leq n < 2.5$ is shown in Figure \ref{alldistq}. 
The peak of the distribution is at $b/a\sim0.4$. 
The $b/a$ distribution of the present-day disks is also shown 
in Figure \ref{alldistq} just to 
demonstrate the $b/a$ distribution in case of a round disk shape. 
The study of a recent, very large ($\sim303,000$) sample of 
disk galaxies from the Sloan Digital Sky Survey Data 
Release 6 (SDSS DR6) by \cite{padilla08} shows a flat $b/a$ distribution 
from $b/a\sim0.2$ to $b/a\sim0.8$. 
It is seen that the $b/a$ distribution of the single-component sBzK galaxies 
is clearly different from that for the local disks; the distribution of the sBzK 
galaxies is skewed toward low $b/a$ with the peak at $b/a \sim 0.4$. 
This implies that the single-component sBzK galaxies do not 
have a round disk structure. 

%--------------------------------------------------------------------------------------Bias on b/a
\subsection{Bias on $b/a$ Distribution}

\begin{figure}
\centering
\includegraphics[clip, angle=-90, width=7cm]{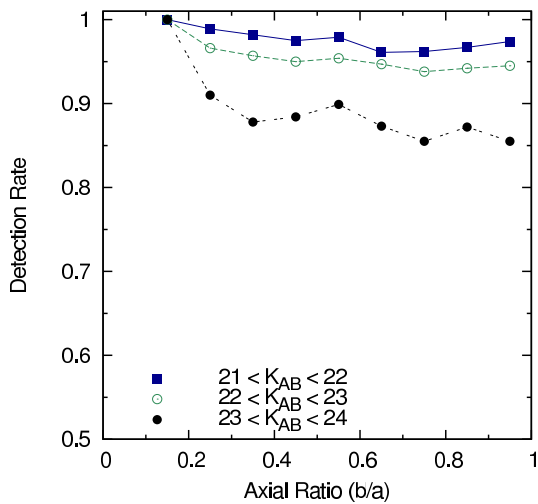}
\caption{
Detection rate of artificial objects in the $K$-band image as a function of 
an apparent axial ratio (b/a). 
Different lines indicate the detection rate of objects in different magnitudes. 
}
\label{detectk}
\end{figure}

Because edge-on galaxies tend to show brighter surface brightness 
than face-on galaxies at the same magnitude, our sample can 
possibly be biased toward the edge-on galaxies at faint magnitudes. 
This may cause the $b/a$ distribution to artificially peak at the lower value. 
In order to test the selection bias in $K$ band, we generated 
10,000 artificial objects with random inclination angles (axial ratios), 
uniformly distributed magnitudes ($21-24$ mag), and random effective 
radii that are comparable to the actual objects observed in the images 
at each magnitude bin. After being convolved with the PSF of the 
$K$-band image, the artificial objects were inserted into the $K$-band 
image and re-detected by using the process identical to that used 
for real galaxies. Figure \ref{detectk} shows the resulting detection rate of 
the artificial galaxies as a function of an input axial ratio. 
For the artificial objects with $K_{\rm AB} = 21-23$ mag, the 
detection rate is almost constant regardless of the axial ratio and 
is not lower than 95\%. 
In the faintest magnitude bin ($23-24$ mag), the detection rate is 
$\sim100$\% at axial ratio ($b/a$) of $0.1-0.2$ (edge-on) and 
decreases to $\sim87\%$ at higher $b/a$ values. 
Though there is a decreasing trend of the detection rate with 
increasing $b/a$, the detection rate remains almost constant 
from $b/a=0.3$ to $b/a=1.0$. Therefore, the selection bias is 
unlikely to be the cause for the peak of the $b/a$ distribution 
of the single-component sBzK galaxies. 

\begin{figure*}
\centering
\begin{tabular}{cc}
\includegraphics[clip, angle=-90, width=7cm]{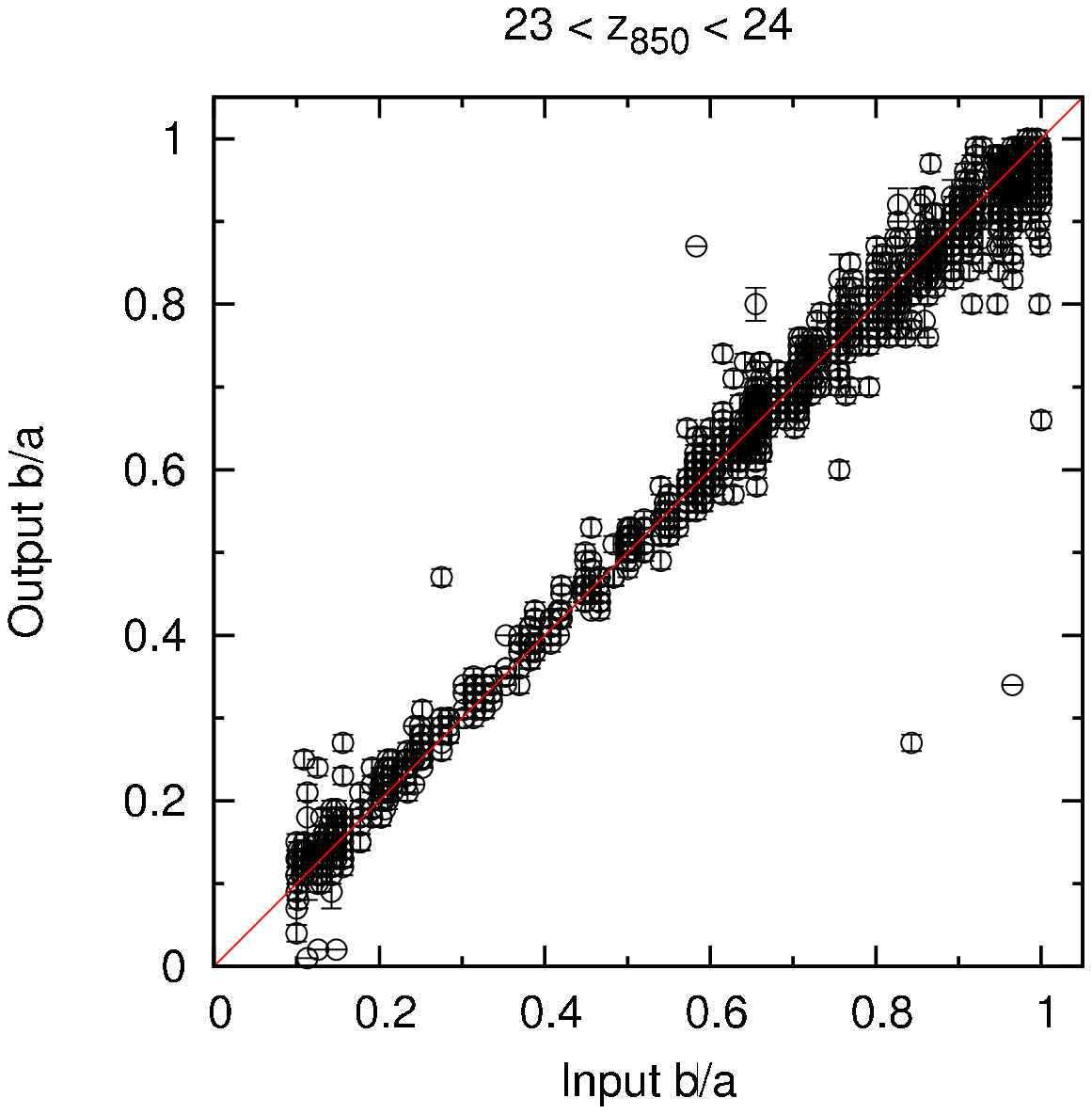} &
\includegraphics[clip, angle=-90, width=7cm]{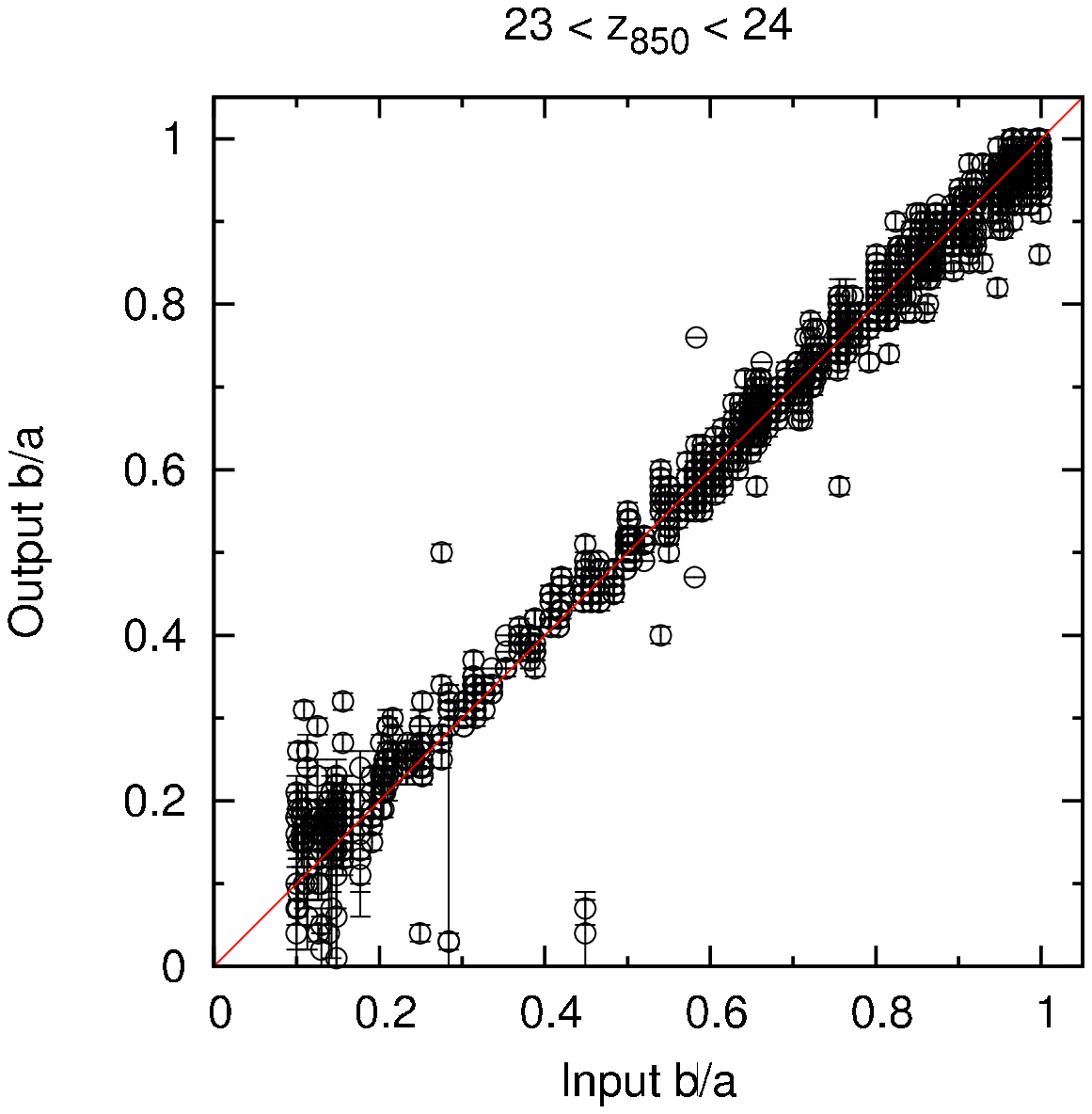} \\
\includegraphics[clip, angle=-90, width=7cm]{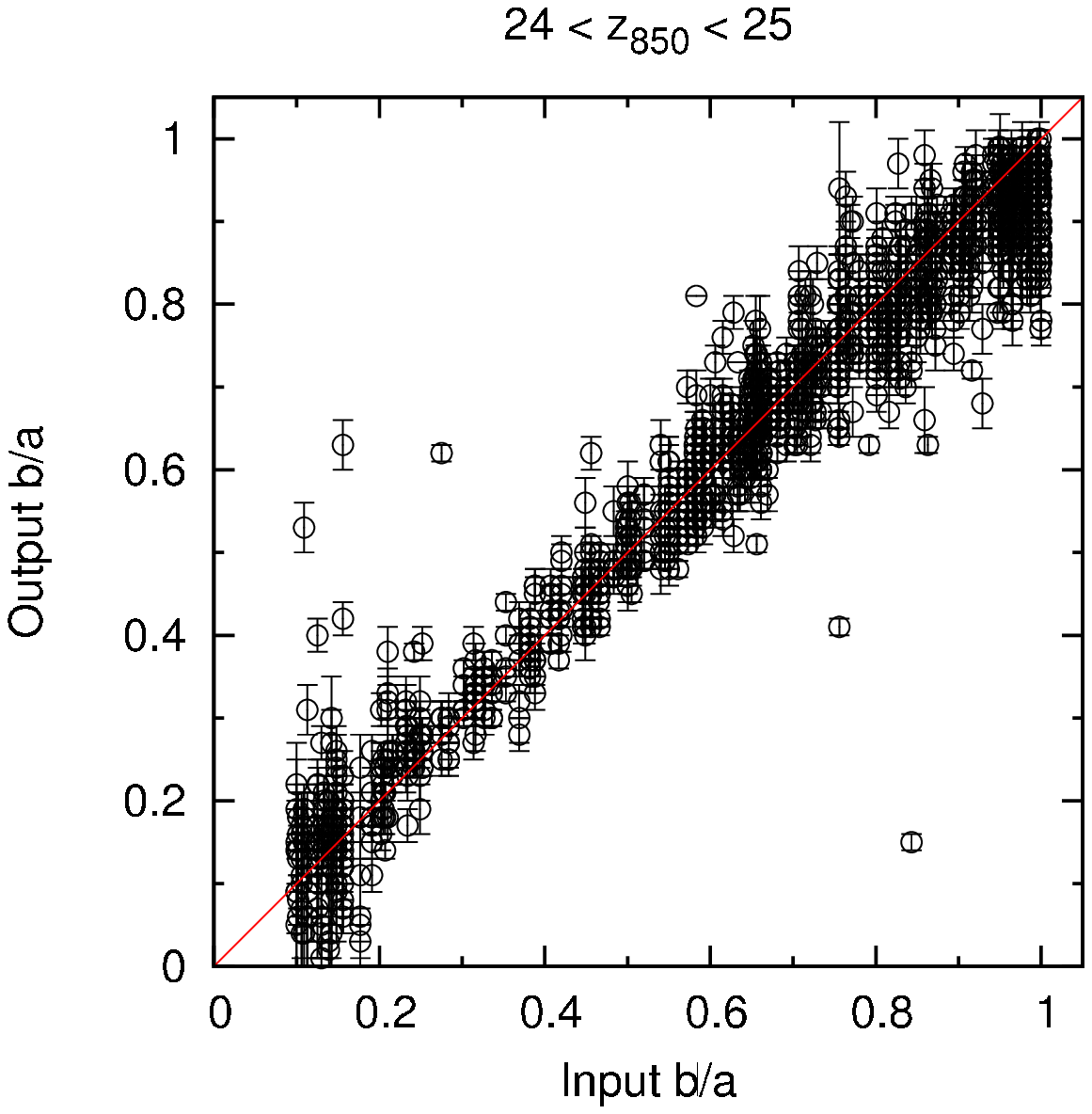} &
\includegraphics[clip, angle=-90, width=7cm]{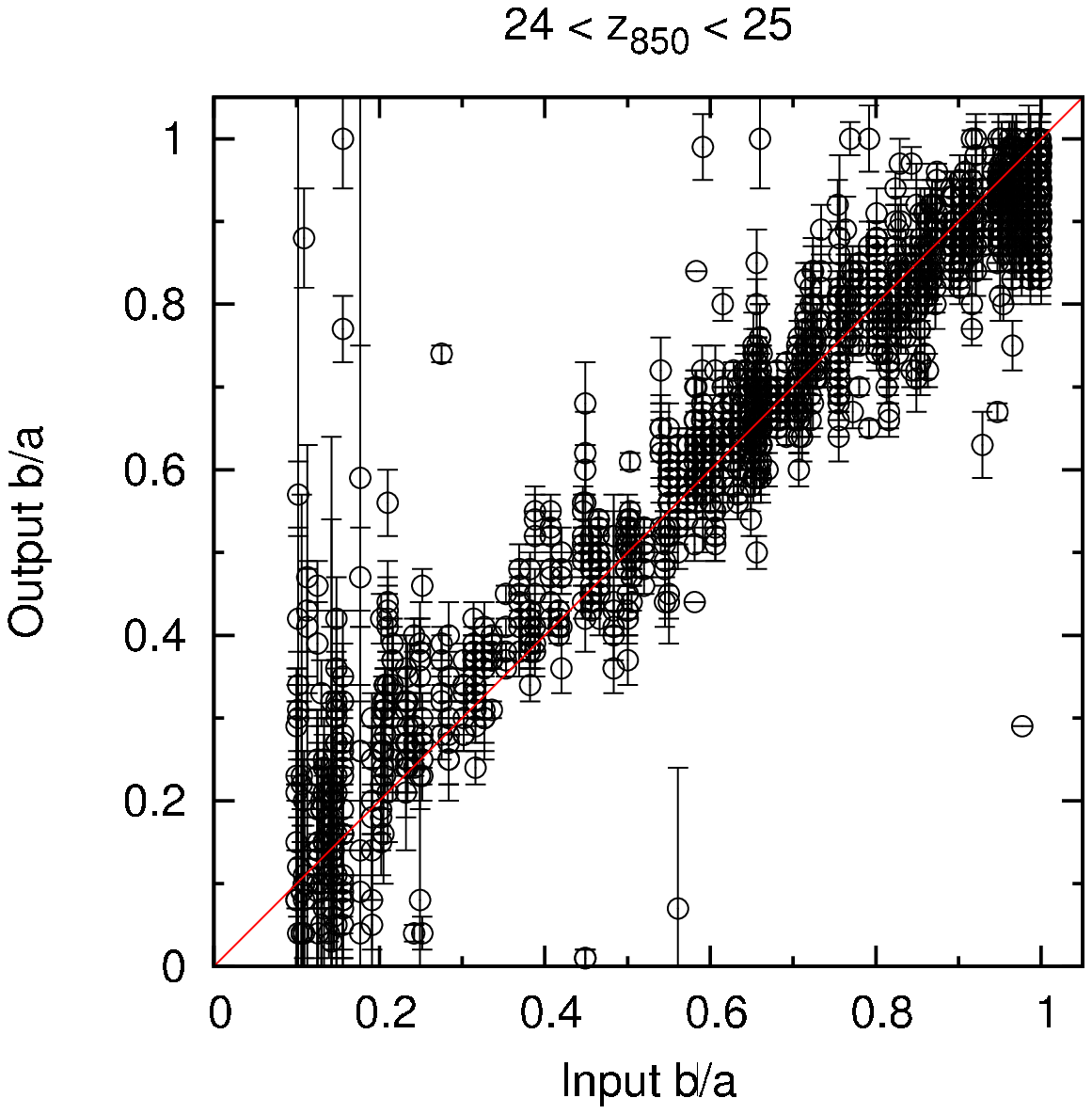} \\
\includegraphics[clip, angle=-90, width=7cm]{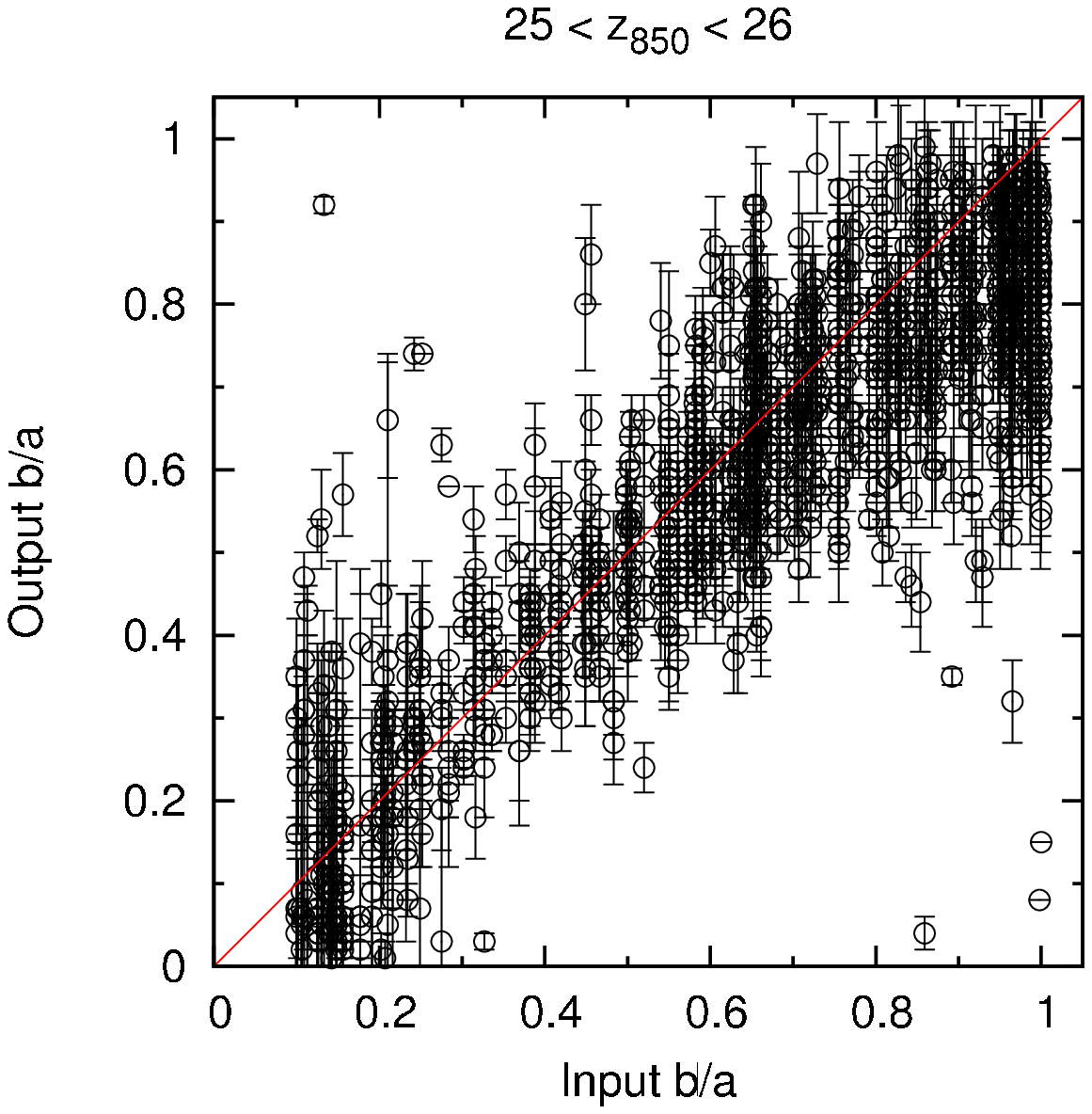} &
\includegraphics[clip, angle=-90, width=7cm]{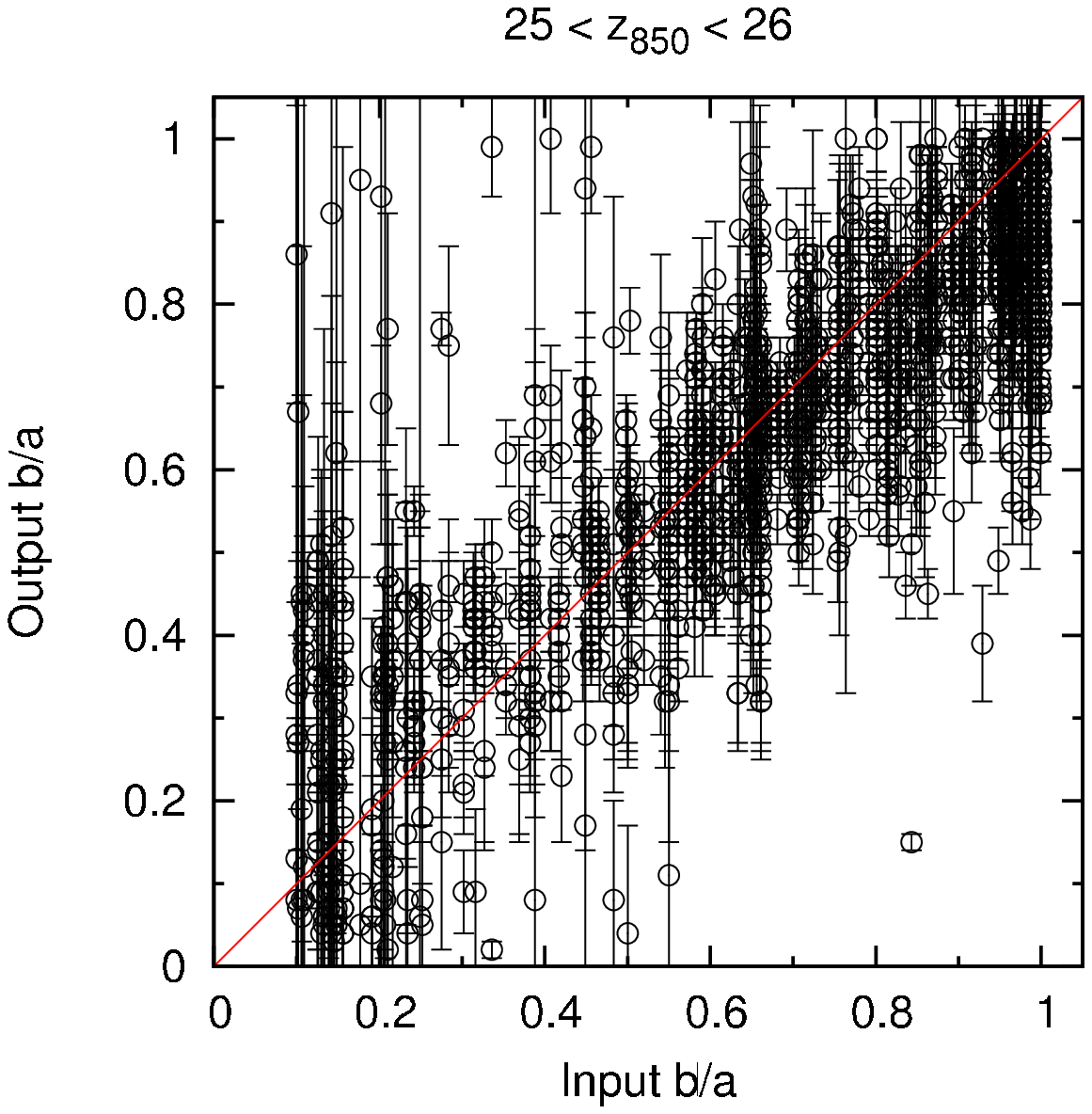} \\
\end{tabular}
\caption{
Comparisons between input and recovered axial ratios ($b/a$) of 
artificial disk objects (left panel) and disk+bulge objects (right panel) 
in \wz~image. 
}
\label{qhist}
\end{figure*}

We also tested this with the \wz~image. 
We generated 5,000 artificial objects with various magnitudes and effective radii 
for each model of an exponential disk and an exponential disk plus 
$r^{1/4}$ bulge (disk+bulge). 
The disk+bulge profile was constructed by assuming the bulge-to-disk 
ratio ($B/D$) of 0.5. The ratio of effective radii of bulge against 
disk is assumed to be $0.1$, which is an average ratio for 
local galaxies in $B$ band \citep{dejong96}. 
The artificial objects were analyzed in the same manner as for 
the sBzK galaxies, i.e., fitting with a single S\'ersic profile. 
The comparisons between the input and the recovered axial ratios 
are shown in Figure \ref{qhist} for disk (left panel) and 
disk+bulge (right panel) profiles. The axial ratios are well 
recovered in $\wz=23-24$ mag as seen in the both models. 
The scatter becomes larger at the fainter magnitudes, 
but no artificial effect which increases the number of output $b/a$ 
around 0.4 is seen in either disk or disk+bulge case. 
Therefore, the peak of the recovered axial 
ratio at $b/a\sim0.4$ seen in our sBzK sample is considered to be real. 

%%%%%%%%%%%%%%%%%%%%%%%%%%%%%%%%%%%%%%%%%
\subsection{Intrinsic Shape of the Single-Component sBzK Galaxies}
\label{subsec:intrinsic}

To examine the intrinsic shape of the sBzK galaxies, we employ
the same method as that used for the local disk galaxies.
 \cite{ryden04} modeled a galaxy as a triaxial ellipsoid 
with major axis $A$, middle axis $B$, and minor axis $C$ 
parameterized by two quantities: 
disk ellipticity ($\epsilon\equiv 1-B/A$) and disk thickness ($C/A$). 
They then fitted the model with the observed distribution of 
apparent axis ratios. 
Following \cite{ryden04}, we adopted a model in which the 
disk ellipticity is assumed to have a lognormal distribution 
with mean $\mu$ ($\equiv \ln \bar{\epsilon}$) and dispersion $\s$, 
while the disk thickness has a Gaussian distribution with 
mean $\mg$ and standard deviation $\sg$. 
These four parameters $\mu$, $\s$, $\mg$ and $\sg$ give 
a distribution of the intrinsic shapes of galaxies. 
By randomly selecting an ellipticity and a disk thickness 
from the distribution,  we can compute the apparent axis ratio 
of $b/a$ \citep{binney85}  at a random viewing angle ($\theta, \phi$)
 using equations  (12)--(15) in \cite{ryden04}. 
Repeating this calculation 5000 times, we obtain 
a model distribution of apparent axial ratio $b/a$, 
which is used to compare with the observed distribution of our sample. 

%Table%%%%%%%%%%%%%%%%%%%%%%%
\begin{deluxetable}{cccc}
\tabletypesize{\footnotesize}
\tablewidth{0pt}
\tablecaption{Grids of Model Parameters\label{model_grid}}
\tablehead{
\colhead{Parameter} & \colhead{Minimum value} & \colhead{Maximum value} & \colhead{Steps}\\
}
\startdata
$\mu$ ($B/A$) & $-4.00$ ($0.98$) & $-0.10$ ($0.01$) & 0.15\\
$\s$ & 0.20 & 2.00 & 0.15\\
$\mg$ & 0.10 & 0.98 & 0.02 \\
$\sg$ & 0.01 & 0.35 & 0.02\\
\enddata
\end{deluxetable}

We constrained the parameter set of $\mu$, $\s$, $\mg$, and $\sg$ 
by fitting with the observed distribution of $b/a$.
Ranges and steps of the parameters examined are listed in Table
\ref{model_grid}. 
The intrinsic $B/A$ ratio corresponding to the $\mu$ parameter
 is also shown in the parenthesis. 
The best-fitting set of parameters was obtained by $\chi^2$
minimization. 
We used a bin size of $\Delta(b/a) = 0.04$ for both the observed 
and model distributions of $b/a$. 
Once we had the best-fitting results, we repeated the fitting analysis 
with a finer grid centered on the best-fitting values. 
This refinement was done twice. 

For the $b/a$ distribution of 318 single-component sBzK galaxies 
with $0.5 \leq n < 2.5$, the final best-fitting parameters are
 $\mu = -0.95^{+0.20}_{-0.15}$, $\s = 0.50^{+0.45}_{-0.15}$,
 $\mg = 0.28^{+0.03}_{-0.04}$, and $\sg = 0.060^{+0.020}_{-0.015}$
 with $\chi^2_\nu = 0.98$. 
The errors are at 68\% confidence level based on Monte Carlo 
realization; we re-derived the best-fitting parameters and 
repeated it 500 times by varying the observed distribution of $b/a$ 
within their Poisson noises. 
The model distribution of the best-fitting parameters is also shown 
with the observed one in Figure \ref{alldistq}. 
The best-fitting $\mu=-0.95$ corresponds to the mean and the peak 
intrinsic $B/A$ ratio of 0.61 and 0.70, respectively. 
It is indicated that the intrinsic shape of the single-component sBzK 
galaxies is more likely to be bar-like or oval, rather than being round. 

\begin{figure*}
\centering 
\begin{tabular}{c c}
\includegraphics[clip, angle = -90, width=7cm]{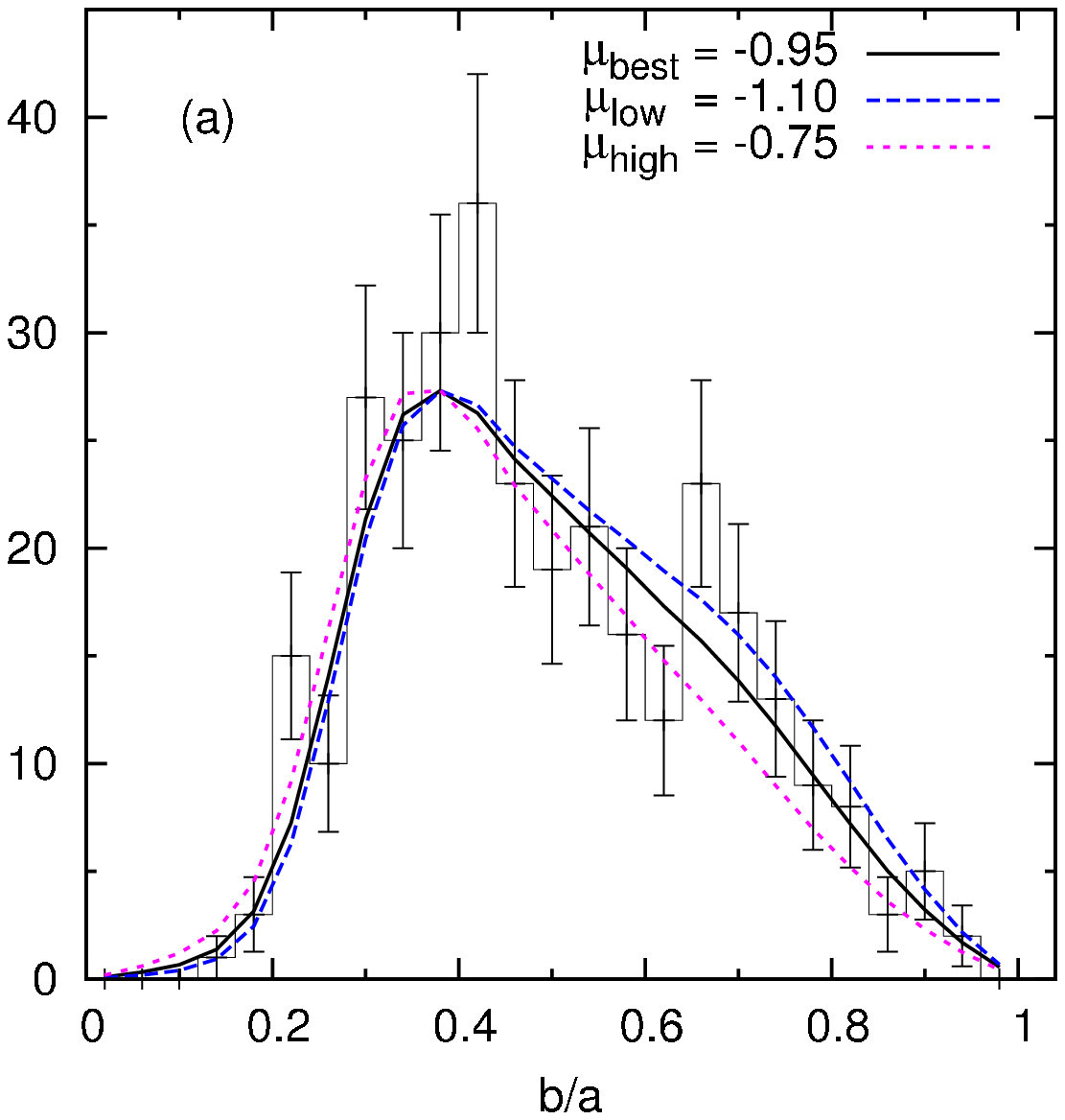} &
\includegraphics[clip, angle = -90, width=7cm]{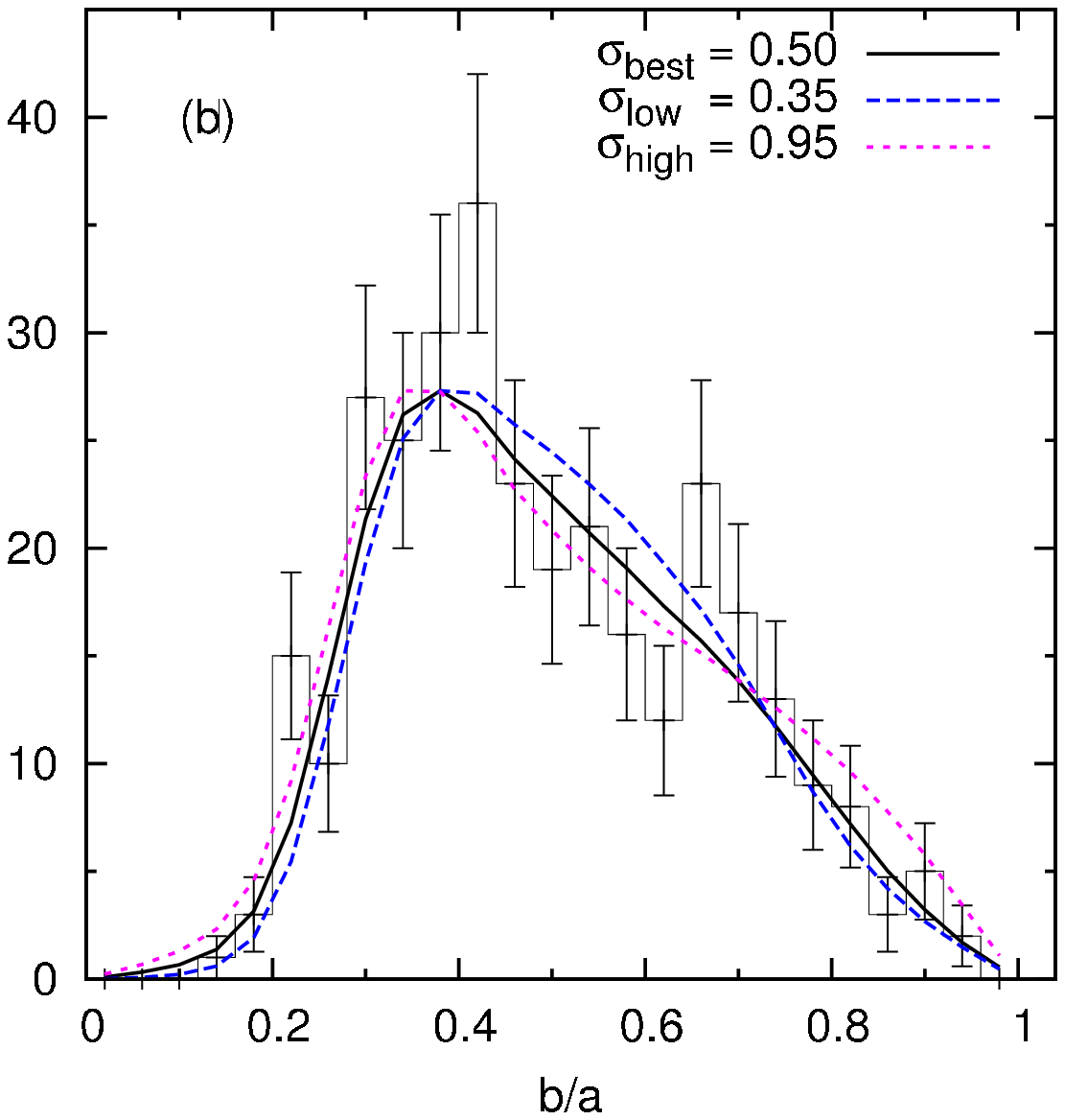} \\
\includegraphics[clip, angle = -90, width=7cm]{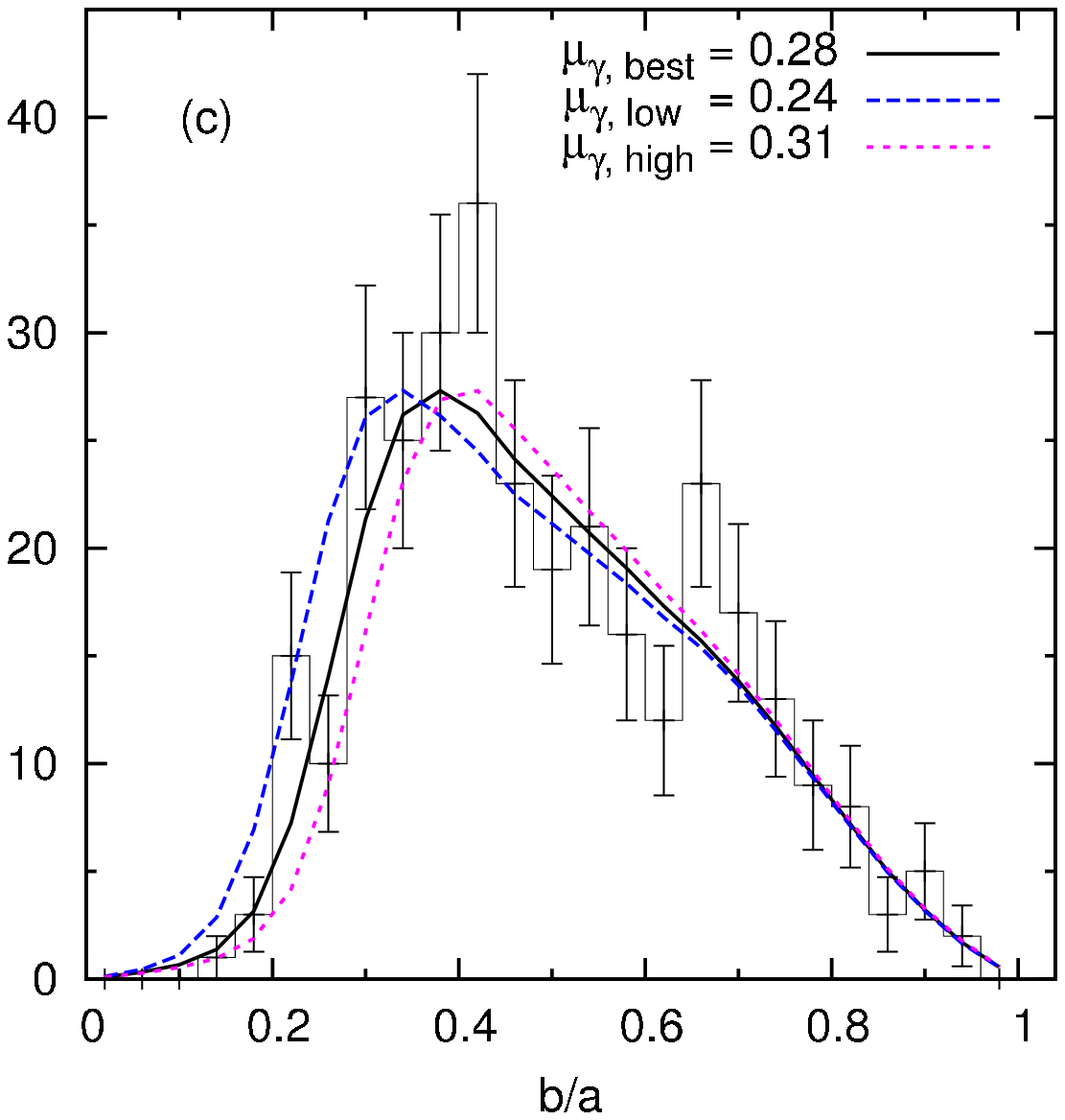} &
\includegraphics[clip, angle = -90, width=7cm]{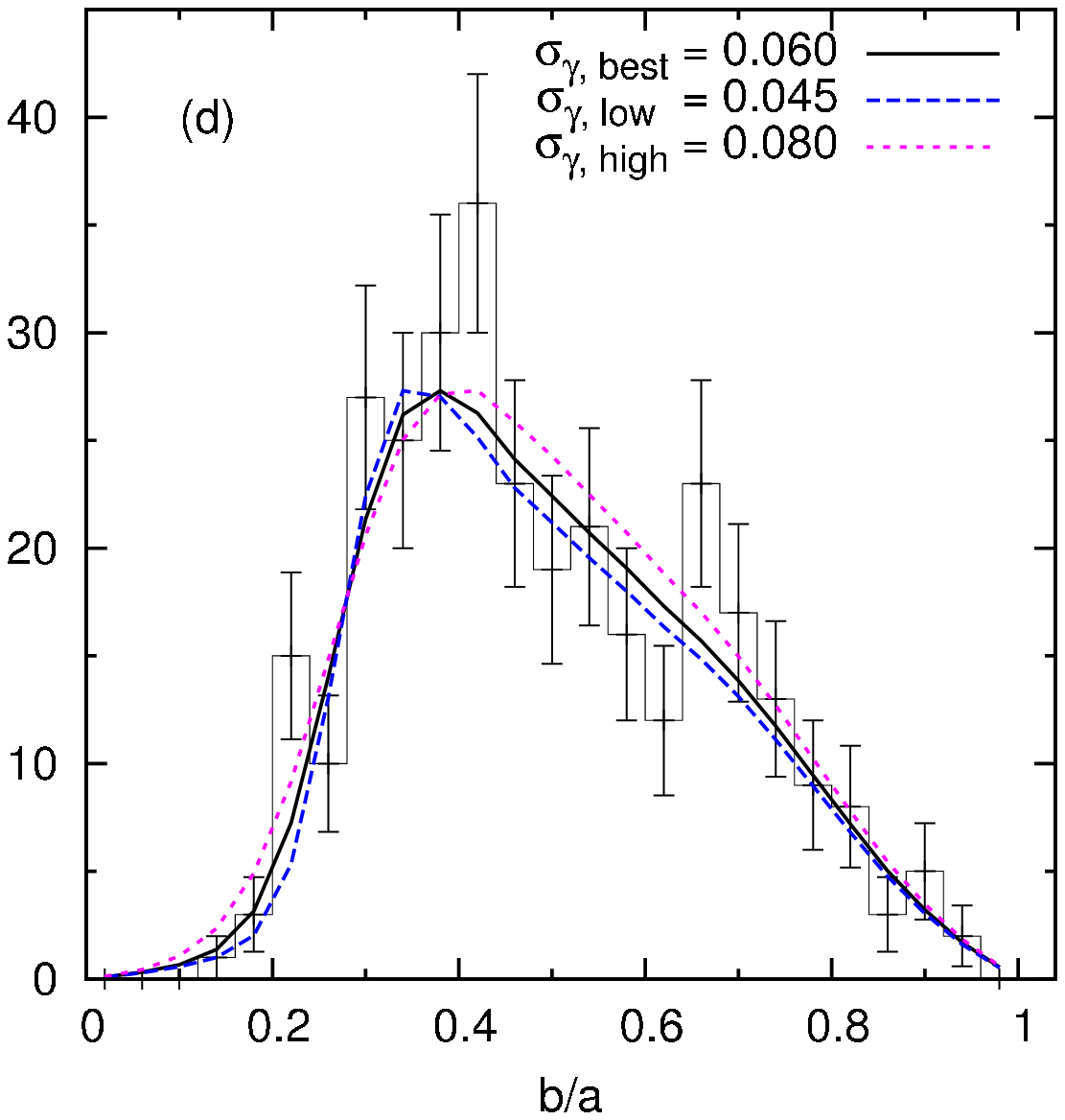} 
\end{tabular}
\caption{Normalized distributions of the apparent axial ratios by varying model parameters. 
From Figures (a) to (d), we vary $\mu$, $\s$, $\mg$, and $\sg$, respectively, 
with 68\% confident interval. 
}
\label{param} 
\end{figure*}

Figure \ref{param}(a) shows the observed distributions of the axial ratios 
together with the best-fitting model and with $1\s$ deviated models 
($\mu = -0.95_{-0.15}^{+0.20}$). The $1\s$ deviated models 
well bracket the observed distribution of apparent axial ratios. 
Figure \ref{param}(a) also shows that the parameter $\mu$
varies the $b/a$ distribution in the right part of the peak;
the frequency of $b/a$ in the part increases with decreasing $\mu$ (i.e.,
increasing $B/A$). 
Changing $\s$ affects the shape of the $b/a$ distribution 
in both sides of the peak, but the effect is not so large (Figure \ref{param}(b)). 
The $\mg$ parameter affects the peak value of the $b/a$ distribution as seen in Figure 
\ref{param}(c). The peak of the apparent axial ratios increases 
with increasing the disk thickness. 
The $\sg$ parameter slightly affects the $b/a$ distribution around the peak 
(Figure \ref{param}(d)). 

\begin{figure*}
\centering
\begin{tabular}{cc}
\includegraphics[angle = -90, width = 5cm]{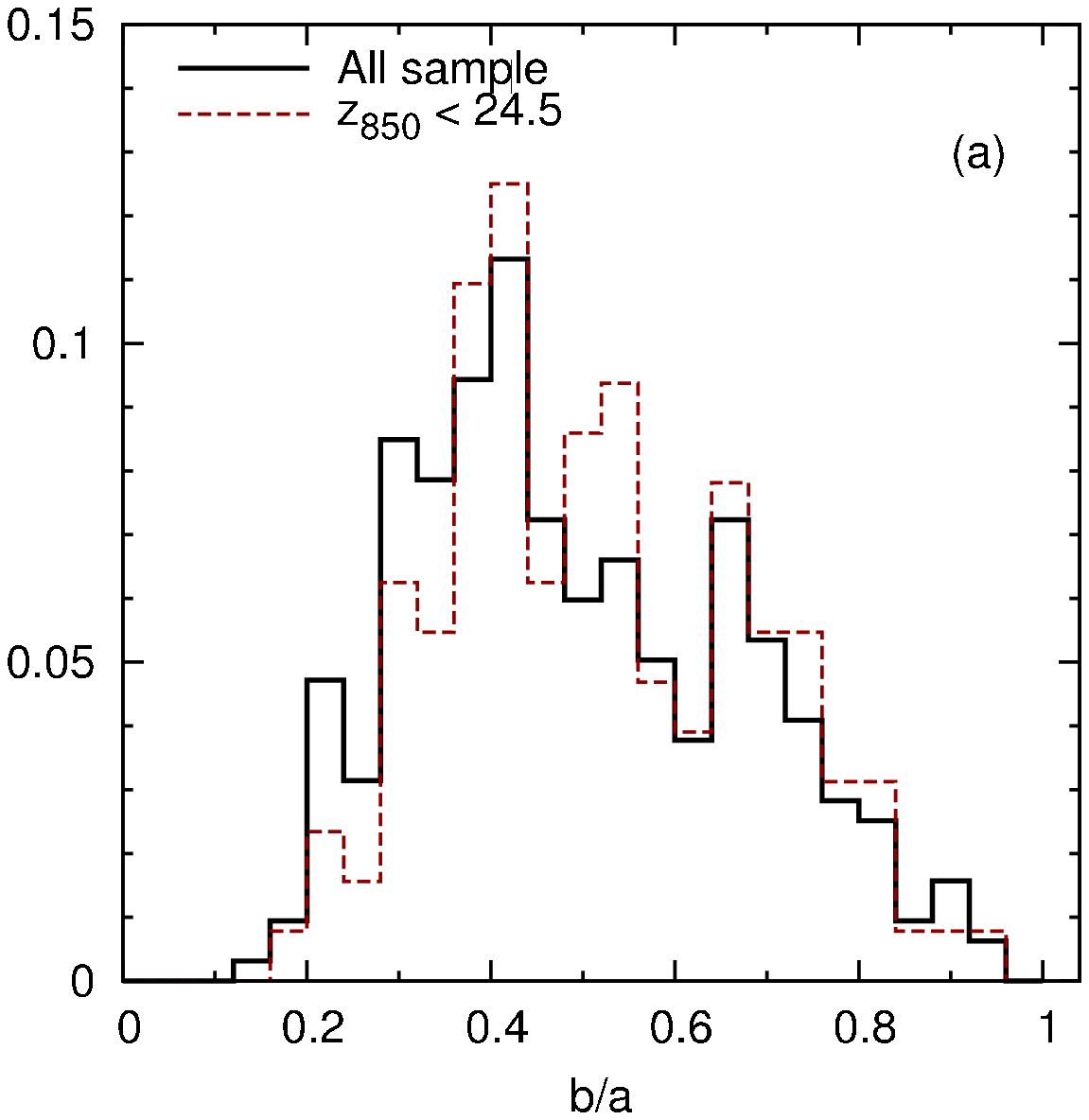}
\includegraphics[angle = -90, width = 5cm]{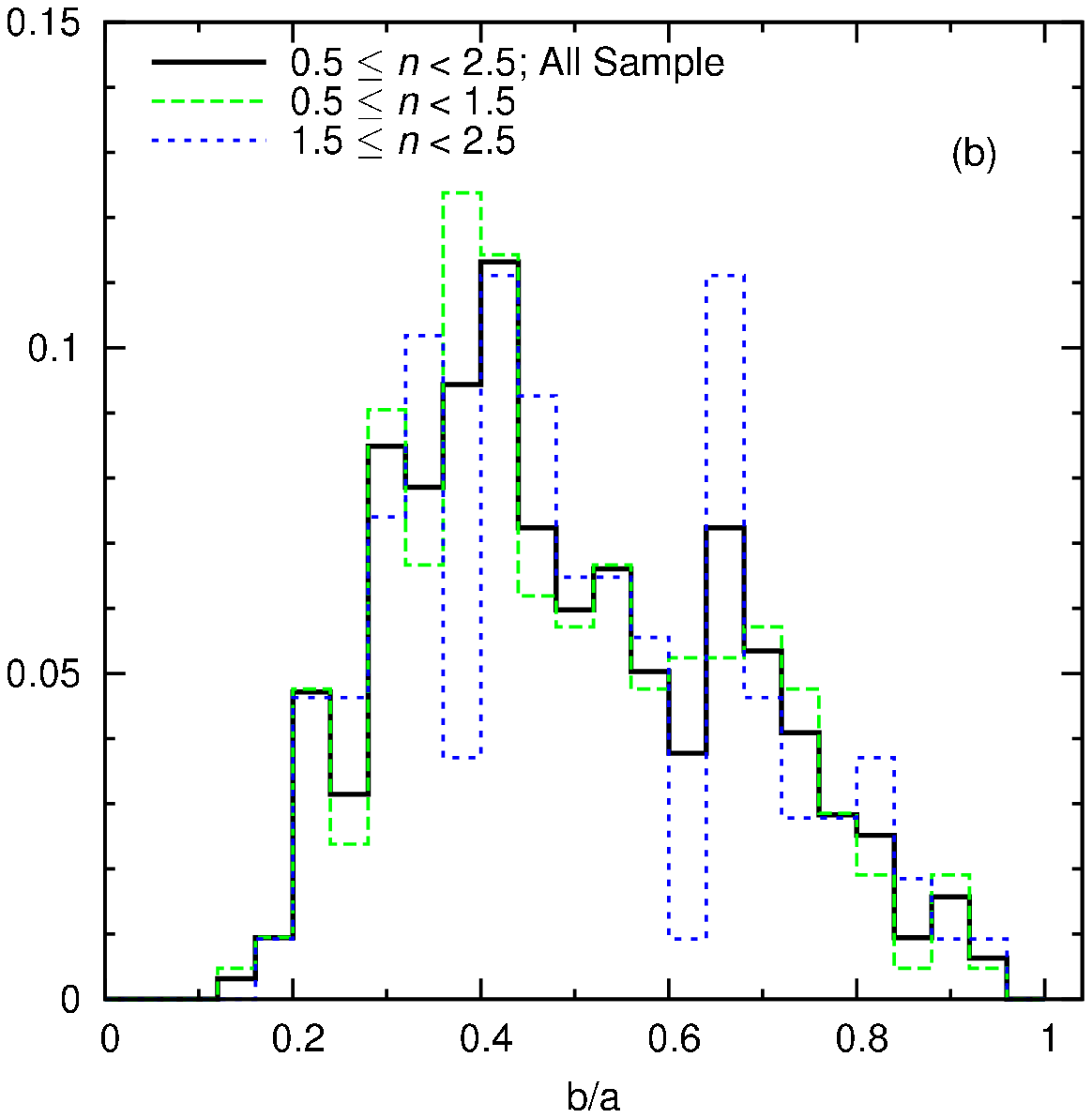} 
\includegraphics[angle = -90, width = 5cm]{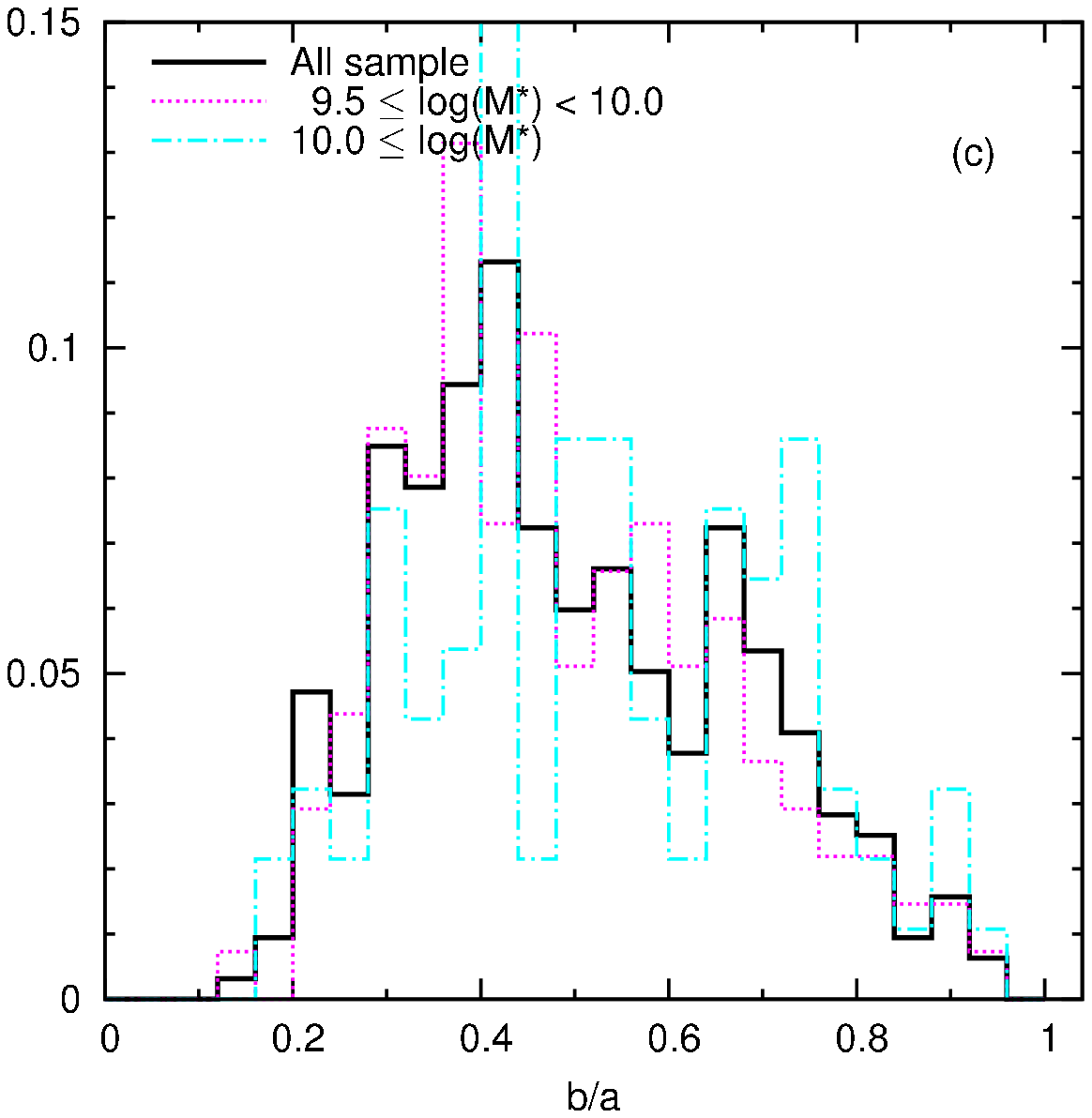} \\ 
\includegraphics[angle = -90, width = 5cm]{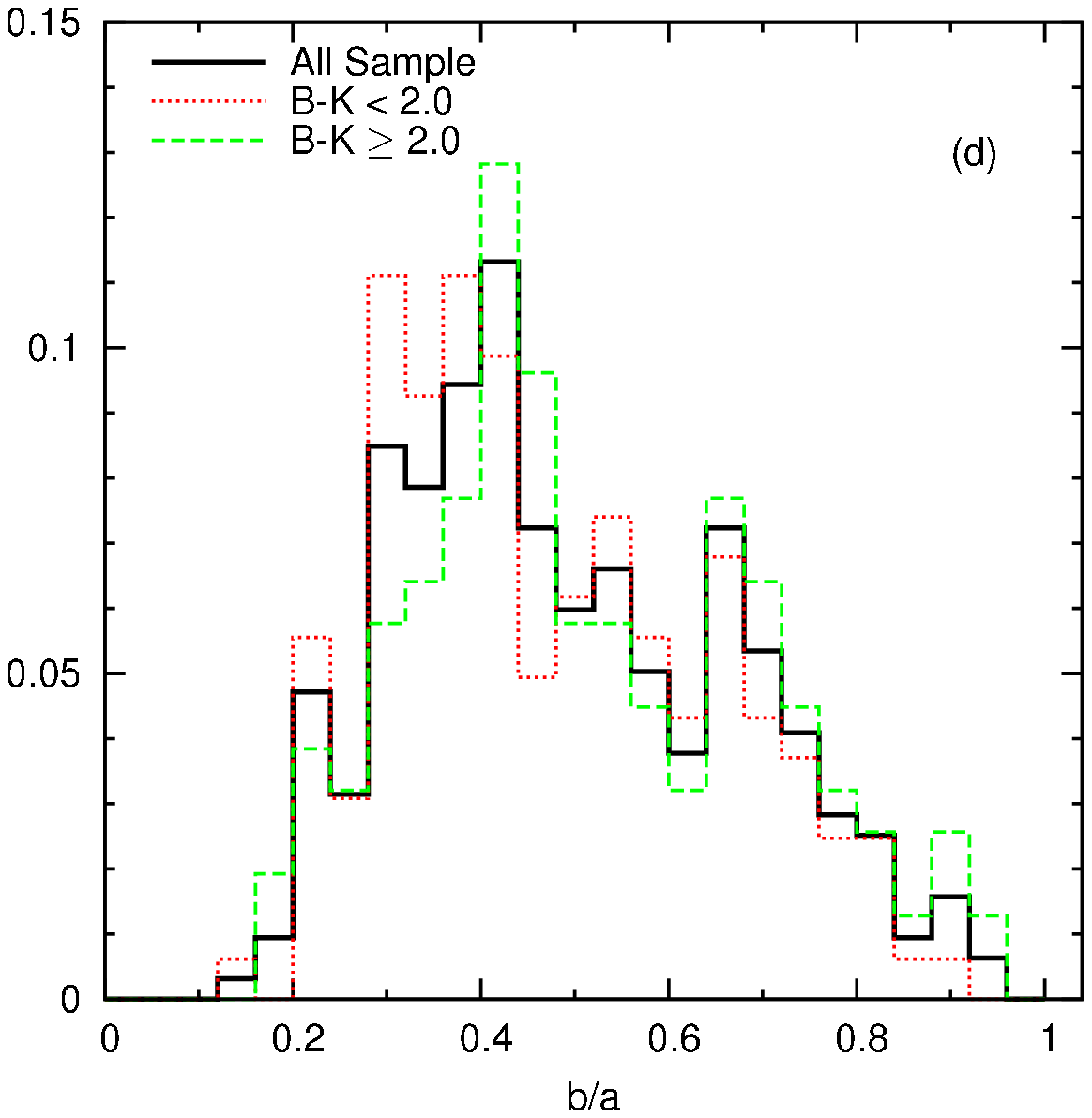} 
\includegraphics[angle = -90, width = 5cm]{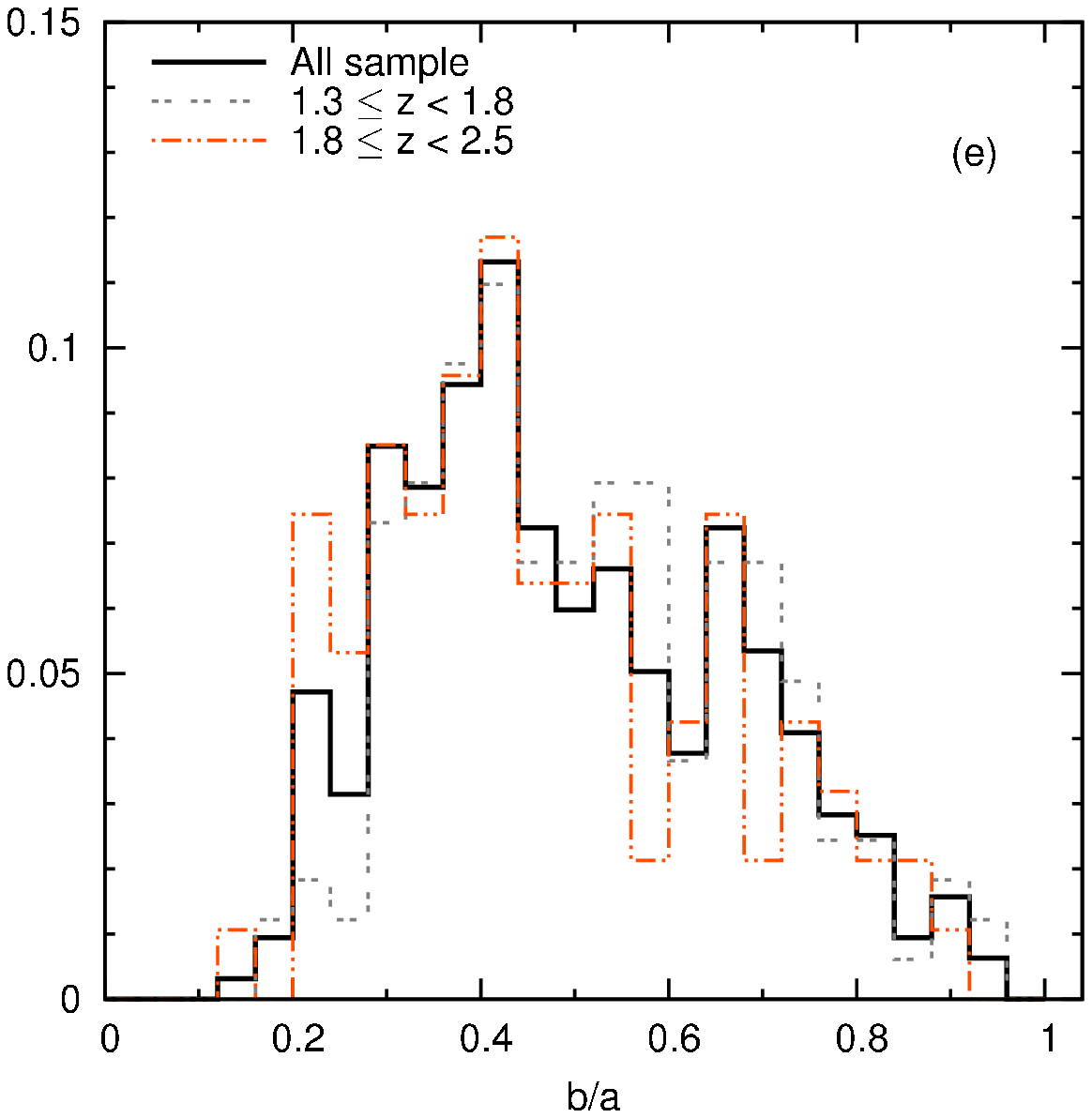} 
\end{tabular}
\caption{Distributions of the apparent axial ratios of sBzK subsamples. 
The whole sample of the single-component sBzK galaxies with $0.5 \leq n < 2.5$ 
is indicated by a solid line in all figures. 
Figure (a) shows the subsample of sBzK galaxies with $\wz <24.5$ mag. 
We divide the galaxies with $0.5 \leq n < 2.5$ into subsamples according 
to their S\'ersic index and their stellar mass in figures (b) and (c), respectively. 
Figure (d) shows the $b/a$ distribution of the subsamples divided by their 
$B-K$ color. The subsamples are divided so that they have approximately 
the same number of galaxies. Figure (e) shows distributions of subsamples 
divided by their photometric redshifts. Ranges of redshifts are selected so 
that both subsamples are in the comparable comoving volume. The 
histograms are normalized so that the area is unity. 
}
\label{dist_subsample} 
\end{figure*} 

We divide the sample of the single-component sBzK galaxies into 
groups according to their properties and show their $b/a$ distributions 
in Figure \ref{dist_subsample}. 
As mentioned above, dividing disk-like and spheroid-like objects at $n=2.5$ is 
efficient down to \wz $<24.5$ mag. 
Because the magnitudes of our sample distribute to $\wz\sim26$ mag, 
we make a subsample of sBzK galaxies with $\wz <24.5$ mag.  
Figure \ref{dist_subsample}(a) shows no significant difference of the $b/a$ histogram 
between the sBzK galaxies with $\wz<24.5$ and $\wz <26.0$ (the whole sample). 
Their best-fitting parameters are also consistent with each other 
as summarized in Table \ref{fitsub}. 
Figure \ref{dist_subsample}(b) shows the distributions for subsamples
 divided by their S\'ersic index ($0.5 \leq n < 1.5$ and $1.5 \leq n < 2.5$). 
The distributions of the subsamples and the whole sample are very similar. 
Their best-fitting parameters agree with those of the whole sample 
almost within $1\s$ (Table \ref{fitsub}). 
Figure \ref{dist_subsample}(c) shows the distributions of the 
apparent $b/a$ for stellar mass subsamples ($10^{9.5} \Msun \leq
M_{\rm stellar} < 10^{10} \Msun$ and $10^{10} \Msun \leq M_{\rm stellar}$). 
The distribution of subsample with $10^{9.5} \Msun \leq 
M_{\rm stellar} < 10^{10} \Msun$ is similar to that of the whole sample, 
while there seems to be a slight difference between the whole sample and 
the objects with $M_{\rm stellar} \geq 10^{10}\Msun$. 
However, the best-fitting parameters listed in Table \ref{fitsub} are 
in agreement mostly within $1\s$ errors, though the errors are 
rather large for some parameters. 
Figure \ref{dist_subsample}(d) illustrates the distributions of the 
sBzK galaxies divided by their $B-K$ color ($B-K < 2.0$ and $2.0 \leq B-K$). 
Both $B-K$ subsamples show the similar distributions to the whole sample. 
Table \ref{fitsub} also indicates that their intrinsic shape 
parameters agree  with those of the whole sample. 
Finally, we divide the sample according to the photometric redshift
($1.3 \leq z_{phot}<1.8$ and $1.8 \leq z_{phot} < 2.5$) 
in Figure \ref{dist_subsample}(e). 
No significant difference between the distributions of subsamples
 and the whole sample is seen from either 
Figure \ref{dist_subsample} or Table \ref{fitsub}, 
suggesting no significant evolution in the redshift range 
which is about one Gyr separation. 

\begin{deluxetable*}{c cccc}
\tabletypesize{\footnotesize}
\tablewidth{0pt}
\tablecaption{Best-Fitting Results of Intrinsic Shape for Subsamples of sBzK Galaxies
\label{fitsub}}
\tablehead{
\colhead{Subsamples} & \colhead{$\mu$} & \colhead{$\s$} & \colhead{$\mg$} & \colhead{$\sg$}\\
}
\startdata
The whole sample & $-0.95^{+0.20}_{-0.15}$ & $0.50^{+0.45}_{-0.15}$ & $0.28^{+0.03}_{-0.04}$ &  $0.060^{+0.020}_{-0.015}$ \\
$\wz <24.5$ & $-0.85^{+0.75}_{-0.45}$ & $0.50^{+0.90}_{-0.30}$ & $0.32^{+0.24}_{-0.06}$ & $0.050^{+0.140}_{-0.040}$ \\
\hline
$0.5 \leq n < 1.5$ & $-0.85^{+0.75}_{-0.30}$ & $0.50^{+0.60}_{-0.30}$ & $0.30^{+0.06}_{-0.06}$ & $0.050^{+0.040}_{-0.040}$\\
$1.5 \leq n < 2.5$ & $-0.85^{+0.75}_{-1.05}$ & $0.95^{+0.90}_{-0.75}$ & $0.26^{+0.26}_{-0.06}$ & $0.050^{+0.240}_{-0.040}$\\
\hline
$ ~9.5 \leq {\rm log}(M^*) < 10.0$ & $-0.85^{+0.75}_{-0.30}$ & $0.50^{+1.05}_{-0.30}$ & $0.28^{+0.06}_{-0.04}$ & $0.050^{+0.040}_{-0.040}$\\
$10.0 \leq {\rm log}(M^*)~~~~~~~~~~~~~$ & $-1.30^{+1.20}_{-0.90}$ & $0.20^{+1.65}_{-0.00}$ & $0.24^{+0.74}_{-0.14}$ & $0.070^{+0.280}_{-0.060}$\\
\hline
$B-K < 2.0$ & $-0.85^{+0.30}_{-0.15}$ & $0.50^{+0.30}_{-0.15}$ & $0.28^{+0.02}_{-0.04}$ & $0.010^{+0.060}_{-0.000}$\\
$B-K > 2.0$ & $-1.00^{+0.90}_{-0.60}$ & $0.80^{+1.05}_{-0.60}$ & $0.30^{+0.34}_{0.08}$ & $0.070^{+0.240}_{-0.040}$\\
\hline
$1.3 \leq z < 1.8$ & $-0.85^{+0.45}_{-0.30}$ & $0.50^{+0.45}_{-0.30}$ & $0.30^{+0.06}_{0.02}$ & $0.010^{+0.080}_{-0.000}$\\
$1.8 \leq z < 2.5$ & $-0.70^{+0.60}_{-0.75}$ & $0.80^{+1.05}_{-0.60}$ & $0.28^{+0.12}_{-0.08}$ & $0.050^{+0.080}_{-0.040}$
\enddata
\end{deluxetable*}

%------------------------------------------------------------------------------------------------------
\section{Possible Origin and Evolution of the Bar-Like Structure}
\label{sec:discussion}
%\subsection{Possible Origin and Evolution of the Bar-Like Structure}

Our analysis of the axial ratio distribution shows that the single-component 
sBzK galaxies have a bar-like or oval structure. 
At this redshift, the outermost isophote of $\mu_{z850}\sim28.0$ 
mag arcsec$^{-2}$ which we reach in this study corresponds to 
$\mu_{B}\sim23.5$ mag arcsec$^{-2}$ at $z\sim0$ by considering 
the cosmological dimming and the average $\wz-J$ color 
of the sample objects. This roughly corresponds to the surface brightness 
just beyond the bar end of present-day galaxies \citep[e.g.,][]{ohta90}. 
Are these direct progenitors of present-day
barred galaxies?
This is unlikely, because the fraction of barred galaxies
decreases with increasing redshifts up to $z\sim1$
(van den Bergh et al. 1996, Abraham et al. 1999, Sheth et al. 2008).
Although there has been a debate on the evolution of the bar fraction
among disk galaxies, a recent study with a large sample taken with HST/ACS 
shows that the bar fraction drops to $\sim 10-20$\% at $z \sim 0.84$
\citep{sheth08}. \cite{sheth08} pointed out in massive galaxies ($\gtrsim 10^{11}
\Msun$) at $z\sim 0.8$ that the fraction is as large as in local massive galaxies, 
but in our sample such massive galaxies are rare and most of the 
sample galaxies have stellar mass less than $10^{11} \Msun$ 
(Figure \ref{photomass}). 
It is also worth noting here that the axial ratio distribution of
nearby barred galaxies is rather flat \citep{ohta90}. 

The radial surface brightness distributions of the single-component 
sBzK galaxies and their intrinsic axial ratios are similar to those of 
N-body bars due to the bar instability. 
The axial ratio ($B/A$) of the N-body bars is $\sim 0.4 - 0.6$ 
and ratio ($C/A$) is $\sim 0.3-0.5$ \citep[e.g.,][]{cs81, ohta90, am02}. 
\cite{am02} examined the morphological differences of bar 
structures formed in various disk models. The shape of the bar 
structure in their massive disk model (MD model), 
where disk dominates in the inner part of a galaxy, 
shows a good similarity with the statistical intrinsic shape of 
the sBzK galaxies. Azimuthally averaged radial profiles of 
the models also show exponential type.
Meanwhile, a halo mass dominated model and a massive disk with a
central bulge model show much more narrower bar structure with a rather 
flat-type radial surface brightness distribution, dissimilar to
the observed properties obtained in this study. 
These suggest that the bar-like structure seen in the sBzK galaxies
may form by bar instability which takes place
when the disk mass fraction against dark halo mass
within a disk radius exceeds the threshold \citep{os73}. 
If this is the case, a key of the origin of the bar-like structure
may be central condensation of baryonic matter as compared with
dark matter distribution; massive baryonic disk in a less massive dark matter
halo within a certain radius can lead  bar instability resulting into
such bar-like structure. 
Although the Q-parameter also plays a role in the bar instability,
the shape of the galaxies may be  a clue to
understand the dynamical formation process of disk galaxies.

Another possible cause for the origin of the bar-like structure
may be galaxy interaction and/or merge \citep[e.g.,][]{no87}.
Since many of the sBzK galaxies show multiple structure (46\%),
this could be  a cause. 
But it should be noted the analyzed sample of the sBzK galaxies
are the single-component sBzK galaxies and thus no clear sign for
strong close galaxy interaction. 
A cold gas accretion onto a galaxy halo \citep[e.g.,][]{dekel09} 
may also form  an non-axisymmetric structure.
However, this kind of cold gas accretion seems to more preferentially
occur at an intersection of filamentary structure of the universe,
and its effect would be dominated in massive galaxy.
\cite{db06} showed that the cold gas accretion
affects in the halo mass larger than $\sim 10^{12} \Msun$ thus
$\sim 10^{11} \Msun$ in baryonic mass, while
the stellar mass of our sample is mostly less than  $\sim 10^{11}
\Msun$. Hence this mechanism would not be expected to be 
dominated in the sBzK galaxies.

The axial ratio distributions peaking at $b/a = 0.4 \sim 0.5$ are
seen among Lyman break galaxies at $z \sim 3$ and 4 \citep{ravindranath06} 
and Lyman $\alpha$ emitters at $z = 3.1$ \citep{gronwall10}. 
Although these higher redshift star-forming galaxies,
in particular Lyman break galaxies, may not be progenitor 
of the present-day disk galaxies as mentioned in section \ref{sec:intro} 
and we see their morphology in UV wavelength, these galaxies 
presumably have similar bar-like structure as the sBzK galaxies, 
and the origin might be the same as that for the sBzK galaxies.

What is the descendant of the sBzK galaxies?
The properties described in section \ref{sec:galfit} and similarity in
size and the stellar mass density point the sBzK galaxies to progenitors 
of the present-day disk galaxies. 
However, since they do not show the round-shape disk structure, 
a transformation from the bar structure to the round disk structure 
should occur, if the sBzK galaxies really evolved into disk galaxies.
The axial ratio distribution of star-forming galaxies at $z\sim 1.2$ is 
rather flat \citep{ravindranath06}, suggesting that galaxies
at $z \sim 1.2$  have the round disk structure.
If the sBzK galaxies are growing into these galaxies at $z\sim1.2$, 
the elapsed time is about $1-2$ Gyr from $z = 1.5 \sim 2$ to $z\sim
1.2$.
Thus a rapid  transformation mechanism is required.
One possibility is that they are gas rich and will make
a gas rich merge and results into an exponential round-shape 
disk \citep[e.g.,][]{springel05}.
Alternatively,  a bulge or a central mass condensation at 
the center of galaxy may dissolve the bar structure.
A growing central mass condensation with a mass of a few to 10 \% of
the disk can dissolve the bar structure with a time scale of a few Gyr
and the bar structure is totally destroyed after $\sim 5$ Gyr for
the MD-type disk \citep{ath05}. 
Thus the growth of a bulge and/or supermassive black hole 
at the center of the galaxies may also cause the shape transformation. 

%%%%%%%%%%%%%%%%%%%%%%%%%%%%%%%%% Conclusion
\section{Conclusion}\label{conclusion} 

We study the intrinsic structure of star-forming BzK galaxies (sBzK) 
at $z\sim2$ in GOODS-North field. 1029 sBzK galaxies were selected 
down to $K_{AB} < 24.0$ mag. 54\% of them shows a single component 
in the ACS/F850LP image, which covers the rest-frame UV 
($\sim3000$\AA) wavelength. Structural parameters of the single-component 
sBzK galaxies were obtained by fitting the two-dimensional light distributions 
in the ACS/F850LP image with a single S\'ersic profile. We found that most 
of them show S\'ersic index of $n=0.5-2.5$, indicative of a disk-like structure. 
The effective radii typically range from 1 kpc to 3 kpc. 
After correcting the effective radii to those in the rest-frame optical 
wavelength, we found in the stellar mass-size diagram that most of the 
single-component sBzK galaxies distribute in the same region as 
$z=0-1$ disk galaxies by \cite{barden05}. This indicates that most of 
the sBzK galaxies show the surface stellar mass density comparable 
to the local and $z\sim1$ disks. 

The peak of S\'ersic-index distribution at $n\sim1$ and the 
comparable surface stellar mass density to the local disks 
suggest that most of the single-component sBzK galaxies 
are disk-like. We further examined their intrinsic shape 
by deriving the distribution of apparent axial ratio ($b/a$) of the 
sBzK galaxies to see whether they are really disk-like galaxies. 
The distribution is skewed toward low $b/a$ values with a peak 
of $b/a\sim0.4$, in contrast to a flat distribution for round-shape disks. 
We compared the axial ratio distribution to the model distributions 
assuming a triaxial ellipsoid model with axes $A>B>C$. 
The best-fitting parameters correspond to the mean 
face-on ratio ($B/A$) of $0.61_{-0.08}^{+0.05}$ and disk 
thickness ($C/A$) of $0.28_{-0.04}^{+0.03}$. 
This indicates that the single-component sBzK galaxies 
have a bar-like shape rather than a round disk shape. 

This bar-like structure is unlikely to be a direct progenitor 
of present-day barred galaxies, since the fraction of 
barred galaxies decreases with increasing redshift. 
The obtained bar structure seems to be similar to 
that formed through bar instability; if it is the case, 
the intrinsic shape may give us a clue to understand dynamical 
evolution of baryonic matter in a dark matter halo. 
If the sBzK galaxies really evolve to the present-day disk galaxies, 
some mechanism for shape transformation is required. 
Gas rich merge or a bulge and/or supermassive black hole 
growth at the center of the galaxies may be responsible 
for this shape transformation.                                                                                                                             

Finally, it is important to note that we studied the intrinsic shape of 
the single-component sBzK galaxies in the rest-frame UV wavelength, 
where structure tends to be influenced by star-forming activities. 
In order to know the distribution of the stellar component,  
further studies in the rest-frame optical wavelength with 
high-resolution images taken with HST/WFC3 is desirable. 
 
We are grateful to the referee for the comments which 
improved the content and clarity of this paper. 
This work is supported by the Grant-in-Aid for Scientific Research on 
Priority Areas (19047003) and the Grant-in-Aid for Global COE program 
"The Next Generation of Physics, Spun from Universality and Emergence" 
from the Ministry of Education, Culture, Sports, Science, and Technology 
(MEXT) of Japan. 
%---------------------------------------------------------------------------------------------------------------


\begin{thebibliography}{}

\bibitem[Abraham et al.(1999)]{abraham99}Abraham, R. G., Merrifield, M. R., Ellis, R. S., Tanvir, N. R., Brinchmann, J. 1999, \mnras, 308, 569
\bibitem[Aguerri et al.(2004)]{aguerri04} Aguerri, J. A. L., Iglesia-Paramo, J., V\'{\i}lchez, J. M., \& Mu\~noz-Tu\~n\'on, C. 2004, \aj, 127, 1344
\bibitem[Akiyama et al.(2008)]{akiyama08} Akiyama, M., Minowa, Y., Kobayashi, N., Ohta, K., Ando, M., \& Iwata, I. 2008, \apjs, 175, 1
\bibitem[Aceves et al.(2006)]{aceves06} Aceves, H., Vel\'azquez, H., \& Cruz, F. 2006, \mnras, 373, 632
\bibitem[Ascaso et al.(2011)]{ascaso11} Ascaso, B., Aguerri, J. A. L., Varela, J., Cava, A., Bettoni, D., Moles, M., \& D'Onofrio, M. 2011, \apj, 726, 69
\bibitem[Athanassoula et al.(2005)]{ath05} Athanassoula, E., Lambert,
  J. C., \&  Dehnen, W. 2005, \mnras, 363, 496
\bibitem[Athanassoula \&  Misiriotis(2002)]{am02} Athanassoula, E., \&
  Misiriotis, A. 2002, \mnras, 330, 35
\bibitem[Barden et al.(2005)]{barden05} Barden, M., et al. 2005, \apj, 635, 959
\bibitem[Barger et al.(2008)]{barger08} Barger, A., J., Cowie, L., L., \& Wang, W. -H. 2008, \apj, 689, 687
\bibitem[Barnes (2002)]{barnes02} Barnes, J. E. 2002, \mnras, 333, 481
\bibitem[Bertin \& Arnouts (1996)]{bertin96} Bertin, E. \& Arnouts, S., 1996, \aap S, 117, 393
\bibitem[Binney(1985)]{binney85} Binney, J. 1985, \mnras, 212, 767
\bibitem[Bolzonella et al.(2000)]{bolzonella00} Bolzonella, M., Miralles, J. -M., Pell\'o, R. 2000, 
\aap, 363, 476
\bibitem[Brinchmann et al. (1998)]{brinch98}  Brinchmann, J., et  al. 1998, \apj, 499, 112
\bibitem[Bruzual \& Charlot (2003)]{bc03} Bruzual, G., \& Charlot, S. 2003, \mnras, 344, 1000 
\bibitem[Bundy et al.(2005)]{bundy05} Bundy, K., Ellis, R. S., Conselice, C. J. 2005, \apj, 625, 621
\bibitem[Calzetti et al. (2000)]{calzetti00} Calzetti, D., Armus, L., Bohlin, C., Kinney, A.L., Koornneef, J., \& Storchi-Bergmann, T. 2000, \apj, 533, 682 
\bibitem[Caon et al. (1993)]{caon93} Caon, N., Capaccioli, M.,  D'Onofrio, M. 1993, \mnras, 265, 1013
\bibitem[Capak et al.(2004)]{capak04} Capak, P., et al. 2004, \aj, 127, 180
\bibitem[Combes \& Sanders (1981)]{cs81} Combes, F., \&  Sanders,
  R. H. 1981, \aap, 96, 164
%\bibitem[Conroy et al.(2008)]{conroy08} Conroy, C., Shapley, A. E., Tinker, J. L., Santos, M. R., \& Lemson, G. 2008, \apj, 679, 1192
\bibitem[Daddi et al.(2004)]{daddi04} Daddi, E., et al. 2004, \apj, 617, 746
\bibitem[Daddi et al.(2005)]{daddi05} Daddi, E., et al 2005, \apj, 626, 680
\bibitem[de Jong(1996)]{dejong96} de Jong, R. S. 1996, \aap, 313, 45
\bibitem[Dekel \& Birnboim (2006)]{db06} Dekel, A., \&  Birnboim,
  Y. 2006, \mnras, 368, 2
\bibitem[Dekel et al. (2009)]{dekel09} Dekel, A., et al. 2009, \nat, 457, 451
\bibitem[D'Onofrio (2001)]{donofrio01} D'Onofrio, M. 2001, \mnras, 326, 1517
\bibitem[Dutton et al.(2011)]{dutton11} Dutton, A. A., et al. 2011, \mnras, 410, 1660
\bibitem[Fall \& Efstathiou (1980)]{fall80} Fall, S. M., \& Efstathiou, G. 1980, \mnras, 193, 189
\bibitem[F\"{o}rster Schreiber et al.(2006)]{forster06} F\"{o}rster Schreiber, N. M., et al. 2006, \apj, 645, 1062
\bibitem[F\"{o}rster Schreiber et al.(2009)]{forster09} F\"{o}rster Schreiber, N. M., et al. 2009, \apj, 706, 1364
\bibitem[Giavalisco \& Dickinson (2001)]{gm01} Giavalisco, M., \&  Dickinson, M. 2001, \apj, 550, 177 
\bibitem[Giavalisco et al.(2004)]{giavalisco04} Giavalisco, M., et al. 2004, \apj, 600, L93
\bibitem[Graham \& Guzman(2003)]{graham03} Graham, A. W., \& Guzm\'an, R. 2003, \aj, 125, 2936
\bibitem[Gronwall et al.(2010)]{gronwall10} Gronwall, C., Bond, N. A., Ciardullo, R., Gawiser, E., Altmann, M., Blanc, G. A., \& Feldmeier, J .J. 2010, arXiv.1005.3006v1
\bibitem[Hartley et al.(2008)]{hartley08} Hartley, W. G., et al. 2008, \mnras, 391, 1301
\bibitem[Hayashi et al.(2007)]{hayashi07} Hayashi, M., Shimasaku, K., Motohara, K., Yoshida, M., Okamura, S., \& Kashikawa, N. 2007, \apj, 660, 72
\bibitem[Ichikawa et al.(2007)]{ichikawa07} Ichikawa, T., et al. 2007, \pasj, 59, 1081
\bibitem[Kajisawa \& Yamada(2001)]{kajisawa01} Kajisawa, M., \& Yamada, T. 2001, \pasj, 53, 833
\bibitem[Kajisawa et al.(2006)]{kajisawa06} Kajisawa, M., et al. 2006, \pasj, 58, 951
\bibitem[Kajisawa et al.(2009)]{kajisawa09} Kajisawa, M., et al. 2009, \apj, 702, 1393
\bibitem[Kong et al.(2006)]{kong06} Kong, X., et al. 2006, \apj, 638, 72
\bibitem[Kormendy et al.(2009)]{kormendy09} Kormendy, J., Fisher, D. B., Cornell, M. E., Bender, R. 2009, \apjs, 182, 216 
\bibitem[Lambas et al.(1992)]{lambas92} Lambas, D. G., Maddox, S. J., \& Loveday, J. 1992, \mnras, 258, 404
\bibitem[Lilly et al.(1998)]{lilly98} Lilly, S., et al. 1998, \apj, 500, 75
\bibitem[Navarro \& Benz (1991)]{navarro91} Navarro, J. F., \& Benz, W. 1991, \apj, 380, 320
\bibitem[Navarro \& Steinmetz(1997)]{navarro97} Navarro, J. F., \& Steinmetz, M. 1997, \apj, 478, 13
\bibitem[Navarro \& White (1994)]{navarro94} Navarro, J. F., \& White, S. D. M. 1994, \mnras, 267, 401
\bibitem[Noguchi (1987)]{no87} Noguchi, M. 1987, \mnras, 228, 635
\bibitem[Ohta et al. (1990)]{ohta90} Ohta, K., Hamabe, M., \& Wakamatsu,
  K. 1990, \apj, 357, 71
\bibitem[Oke et al.(1983)]{oke83} Oke, J. B., \& Gunn, J. E. 1983, \apj, 266, 713
\bibitem[Ostriker \& Peebles (1973)]{os73} Ostriker, J. P., \&
  Peebles, P. J. E. 1973, \apj, 186, 467
\bibitem[Ouchi et al. (2001)]{ouchi01} Ouchi, M., et al. 2001, \apj,  558, L83
\bibitem[Overzier et al.(2010)]{overzier10} Overzier, R. A., Heckman, T. M., Schiminovich, D., Basu-Zych, A., Gon\c calves, T., Martin, D. C., \& Rich, R. M. 2010, \apj, 710, 979
\bibitem[Padilla \& Strauss(2008)]{padilla08} Padilla, N. D., \&
  Strauss, M. A. 2008, \mnras, 388, 1321
\bibitem[Pannella, et al. (2006)]{pannella06} Pannella, M., Hopp, U.,  Saglia, R. P., Bender, R., Drory, N., Salvato, M., Gabasch, A.,  Feulner, G. 2006, \apj, 639, L1
\bibitem[Peng et al.(2002)]{peng02} Peng, C. Y., Ho, L. C., Impey, C. D., \& Rix, H.-W. 2002, \aj, 124, 266
\bibitem[Pickles (1998)]{pickles98} Pickles, A. J. 1998, \pasp, 110, 863
\bibitem[Ravindranath et al.(2006)]{ravindranath06} Ravindranath, S., et al. 2006, \apj, 652, 963
\bibitem[Robertson et al.(2006)]{robertson06} Robertson, B, Bullock, J. S., Cox, T. J., Di Matteo, T, Hernquist, L, Springel, V, \& Yoshida, Naoki 2006, \apj, 645, 986
\bibitem[Rothberg \& Joseph (2004)]{rothberg04} Rothberg, B., \& Joseph, R. D., 2004, \aj, 128, 2098
\bibitem[Ryden(2004)]{ryden04} Ryden, B. S. 2004, \apj, 601, 214
\bibitem[Sales et al. (2010)]{sales10} Sales, L. V., Navarro, J. F., Schaye, J., Vecchia, C. D., Springel, V., \& Booth, C. M., 2010, \mnras, 409, 1541
\bibitem[Salpeter (1955)]{salpeter55} Salpeter, E. E. 1955, \apj, 121, 161
\bibitem[Sargent et al.(2007)]{sargent07} Sargent, M. T., et al. 2007,  \apjs, 172, 434
\bibitem[Scarlata et al. (2007)]{sca} Scarlata, C., et al.  2007,  \apjs, 172, 406
\bibitem[S\'ersic(1963)]{sersic63} S\'ersic, J. L. 1963, BAAA, 6, 41
\bibitem[S\'ersic(1968)]{sersic68} S\'ersic, J. L. 1968, Atlas de Galaxias Australes (C\'ordoba: Obs. Astron., Univ. Nac. C\'ordoba)
\bibitem[Shen et al.(2003)]{shen03} Shen, S., Mo, H. J., White, S. D. M., Blanton, M. R., Kauffmann, G., Voges, W., Brinkmann, J., \& Csabai, I. 2003, \mnras, 343, 978
\bibitem[Sheth et al.(2008)]{sheth08} Sheth, K., et al. 2008, \apj, 675, 1141
\bibitem[Springel \& Hernquist (2005)]{springel05} Springel, V., \& Hernquist, L. 2005, \apj, 622, L9
\bibitem[Steidel et al. (1996)]{steidel96} Steidel, C. C., Giavalisco, M., Pettini, M., Dickinson, M., Adelberger, K. L. 1996, \apj, 462, L17
\bibitem[Suzuki et al.(2008)]{suzuki08} Suzuki, R., et al. 2008, \pasj, 60, 1347
\bibitem[Swinbank et al.(2010)]{swinbank10} Swinbank, A. M., et al. 2010, \mnras, 405, 234
\bibitem[Unterborn \& Ryden(2008)]{unterborn08} Unterborn, C. T., \& Ryden, B. S. 2008, 
\apj, 687, 976
\bibitem[van den Bergh et al. (1996)]{berg96} van den Bergh, S.,
  Abraham, R. G., Ellis, R. S., Tanvir, N. R., Santiago, B. X.,
  \& Glazebrook, K. G. 1996, \aj, 112, 359 
\bibitem[Vincent \& Ryden(2005)]{vincent05} Vincent, R. A., \& Ryden, B. S. 2005, \apj, 623, 137
\bibitem[Williams et al.(1996)]{williams96} Williams, R. E., et al. 1996, \aj, 112, 1335
\bibitem[Wirth et al.(2004)]{wirth04} Wirth, G. D, et al. 2004, \apj, 127, 3121
\bibitem[White \& Rees (1978)]{white78} White, S. D. M., \& Rees, M. J. 1978, \mnras, 183, 341
\bibitem[Yoshikawa et al.(2010)]{yoshikawa10} Yoshikawa, T., et al. 2010, \apj, 718, 112
\bibitem[Yuma et al.(2010)]{yuma10} Yuma, S., Ohta, K., Yabe, K., Shimasaku, K., Yoshida, M., Ouchi, M., Iwata, I., \& Sawicki, M. 2010, \apj, 720, 1016

\end{thebibliography}
\end{document}